\documentclass[review]{elsearticle}

\usepackage[utf8]{inputenc}
\usepackage[T1]{fontenc}
\usepackage{tgtermes}
\usepackage{lineno,hyperref,pifont,natbib,geometry,fleqn,graphicx,makeidx,fancyhdr,multirow,verbatim,amsmath,amsthm,amssymb,float,array,subfloat,epstopdf,subfig,upgreek,appendix,tabularx}
\modulolinenumbers[5]

\journal{Journal of Instrumentation}

\makeatletter
\def\ps@pprintTitle{%
\let\@oddhead\@empty
\let\@evenhead\@empty
\def\@oddfoot{}%
\let\@evenfoot\@oddfoot}
\makeatother

\begin{document}

\begin{frontmatter}

\title{Charge Collection and Electrical Characterization 
of Neutron Irradiated Silicon Pad Detectors for the CMS High Granularity Calorimeter} 


\author[a]{N. Akchurin}
\author[f]{P. Almeida}
\author[d]{G. Altopp}
\author[e]{M. Alyari}
\author[g]{T. Bergauer}
\author[f]{E. Brondolin}
\author[d]{B. Burkle}
\author[c]{W. D. Frey}
\author[e]{Z. Gecse}
\author[d]{U. Heintz}
\author[d]{N. Hinton}
\author[b]{V. Kuryatkov}
\author[e]{R. Lipton}
\author[f]{M. Mannelli}
\author[a]{T. Mengke}
\author[g]{P. Paulitsch}
\author[a]{T. Peltola\corref{cor2}}
\ead{timo.peltola@ttu.edu}
\author[g]{F. Pitters}
\author[f]{E. Sicking}
\author[d]{E. Spencer}
\author[r]{M. Tripathi}
\author[f]{M. Vicente Barreto Pinto}
\author[d]{J. Voelker}
\author[a]{Z. Wang}
\author[h]{R. Yohay}

\address[a]{Texas Tech University, Department of Physics and Astronomy, Lubbock, TX 79409, USA}
\address[f]{CERN, 1211 Geneva 23, Switzerland}
\address[d]{Brown University, Department of Physics, Providence, RI 02912, USA}
\address[e]{Fermi National Accelerator Laboratory, PO Box 500, Batavia, IL 60510, USA}
\address[g]{Institut f{\"u}r Hochenergiephysik, 1050 Vienna, Austria}
\address[c]{McClellan Nuclear Reactor Center, UC Davis, Davis, CA 95616, USA}
\address[b]{Texas Tech University, Nanotech Center, Lubbock, TX 79409, USA}
\address[r]{University of California Davis, Physics Department, Davis, CA 95616, USA}
\address[h]{Florida State University, Tallahassee, FL 32306, USA}



\cortext[cor2]{Corresponding author}




\begin{abstract}
\label{Abstract}

%
The replacement of the existing endcap 
calorimeter in the Compact Muon Solenoid (CMS) detector for the high-luminosity 
LHC (HL-LHC), scheduled for 2027, will be a high granularity calorimeter.
It will provide 
detailed position, energy, and timing information on electromagnetic and hadronic showers in the immense pileup of the HL-LHC. 
The High Granularity Calorimeter
(HGCAL) 
will use 120-, 200-, and 300-$\upmu\textrm{m}$ thick 
silicon (Si) pad sensors 
as the main active material 
and will 
sustain 
1-MeV neutron equivalent fluences 
up to about $10^{16}~\textrm{n}_\textrm{eq}\textrm{cm}^{-2}$. 
%
%
%
In order 
to address 
the performance degradation of the Si detectors 
caused by the intense radiation environment, 
irradiation campaigns of test diode samples from 
8-inch and 6-inch wafers were 
performed in two 
reactors. 
Characterization of the electrical and charge collection properties after irradiation 
involved both bulk polarities for the three sensor thicknesses. 
Since the Si sensors 
will be operated at -30 $^{\circ}$C to reduce increasing bulk leakage current with fluence, the charge collection investigation of 30 irradiated samples was carried out with the infrared-TCT setup at -30 $^{\circ}$C. 
TCAD simulation results at 
the lower fluences 
are in close agreement with the experimental results and 
provide predictions of sensor performance for the lower fluence regions not covered by the experimental study. All investigated sensors display 60\% or higher charge collection efficiency at their respective highest lifetime fluences 
when operated at 800 V, and display above 90\% 
at the lowest fluence, 
at 600 V. The collected charge close to the fluence of $10^{16}~\textrm{n}_\textrm{eq}\textrm{cm}^{-2}$ exceeds 1 fC at voltages beyond 800 V. 

\end{abstract}

%
\end{frontmatter}


\newpage
\section{Introduction}
\label{Introduction}
Silicon (Si)-based high granularity technology was chosen 
for replacement of the existing 
CMS experiment's endcap calorimeter for the era of the high-luminosity 
Large Hadron Collider (HL-LHC) \cite{Gianotti2005}. The 
HL-LHC is
expected to increase the instantaneous luminosity of LHC by 
a factor of five and deliver an estimated integrated luminosity 
up to 4,000~fb$^{-1}$ 
in 10 years of
operation\footnote{https://home.cern/resources/faqs/high-luminosity-lhc}, 
resulting in maximum
integrated doses in excess of 1.5~MGy and neutron fluences of about 
$1.0\times10^{16}~\textrm{n}_\textrm{eq}\textrm{cm}^{-2}$ 
that the 
CMS endcap calorimeters 
must sustain~\cite{Phase2}. 

The 
High Granularity Calorimeter (HGCAL) 
aims to perform 3D-position, energy and time 
measurements 
in the endcap region 
($1.5 \leq |\eta| \leq 3.2$). 
The 
electromagnetic calorimeter (CE-E) consists of 28 layers of silicon sensors, with Cu, Cu/W, and Pb absorbers. 
Similar silicon sensor technology is utilized in the 
high-radiation region of the hadronic calorimeter (CE-H) 
for the 
8 front layers interleaved with steel absorber plates, while the following 14 layers 
combine silicon sensors, 
with plastic scintillator readout by silicon photomultipliers (SiPM) in the higher- and lower-$\eta$ regions, respectively. 
The approximately 27,000 hexagonal 8-inch silicon modules\footnote{The prototyping was done with sensors originating from 6-inch wafers, while 8-inch wafers were chosen to reduce the total number of modules needed for the HGCAL.} 
represent about 6M channels and an area 
of 600 m$^2$. The hexagonal shape of the sensors
maximizes the usable area of the circular wafers 
while 
remaining tileable. 
The main properties of the 
silicon 
detector channels, 
along with their expected fluences 
for the HGCAL, are given in Table~\ref{tabCells}. 
The detectors must operate in a high-radiation environment for 
a decade, while maintaining sufficient signal-to-noise performance for minimum-ionizing particle (MIP) detection. Clean MIP detection in each sensor cell is required to achieve acceptable inter-cell calibration accuracy. 
Macroscopic radiation damage in the silicon bulk manifests 
as increase in 
leakage current, space
charge, and trapping of the drifting charge, 
all of which are detrimental to 
the performance of a silicon detector 
\cite{Fretw2007,Affolder2011}.

The increase of leakage current with fluence will be 
mitigated by operating the silicon sensors at -30 $^{\circ}$C (243 K). In response to the trapping of the drifting charge, 
thinner sensors will be 
used at the 
higher-$\eta$ 
(smaller radius) regions of the endcaps, 
along with high voltage operation ($\geq$ 600 V 
for heavily-irradiated sensors) and 
readout at segmented $\textrm{n}^{+}$ electrodes collecting electrons ($n$-on-$p$ sensors) in all regions. 
In this study, we discuss experimental and simulation results 
from 
the test diodes diced from the full 
sensor wafers. 

\begin{table*}[!t]
\centering
\caption{Hexagonal Si sensor cells (DC-coupled planar diodes without biasing structure) in CE-E and CE-H all-silicon layers, showing 
the properties of different sensor types 
and the expected 
1-MeV neutron equivalent fluence for each type 
after an estimated integrated luminosity of
4,000 fb$^{-1}$. 
The upper fluence limits are the current estimates for the highest levels that may be reached, which are somewhat larger than in \cite{Phase2}.
%
The lowest lifetime fluences at the outer radii of each region are roughly equal to the highest fluence of the next region, while 
in the case of 300-$\upmu\textrm{m}$ thick sensors, 
it is about $1\times10^{14}~\textrm{n}_\textrm{eq}\textrm{cm}^{-2}$. 
%
} 
\label{tabCells}
\begin{tabular}{|c|c|c|c|}
    \hline
    {\bf Active thickness} [$\upmu\textrm{m}$] & 300 & 200 & 120\\
    \hline
     {\bf Cell size} [$\textrm{cm}^{2}$] & 1.18 & 1.18 & 0.52\\
    \hline
     {\bf Cell capacitance} [pF] & 45 & 65 & 50\\
    \hline
     {\bf Highest lifetime fluence ($\Phi_\textrm{max}$)} [$\textrm{n}_\textrm{eq}\textrm{cm}^{-2}$] & $(5-6)\times10^{14}$ & $(2.5-3)\times10^{15}$ & $(0.92-1)\times10^{16}$\\
    \hline
\end{tabular}
\end{table*}
The studies of the detector 
response, 
which used a sub-nanosecond IR-laser, 
provide information on the carrier transport 
and electric field distribution in the sensor 
bulk, which are 
the input 
parameters for the 
long-term prediction 
of the detector's 
charge collection efficiency (CCE) 
with increasing fluence. 
By complementing observed CCE with leakage current and capacitance measurements, 
an extensive picture of the macroscopic effects from
the microscopic radiation-induced defects in the silicon bulk 
is obtained that 
continue to refine guidelines for detector design and operation. Since the 
fluence at HGCAL is dominated by 
neutrons \cite{Phase2}, the 
radiation hardness studies were carried out with reactor neutrons.  

In a previous 
study 
\cite{Curras2017} 
with neutron irradiated deep-diffused float-zone (dd-FZ) and epitaxial test diodes at -20 $^{\circ}$C, 
the 300-$\upmu\textrm{m}$-thick $n$-types 
were found to 
collect significantly 
more charge than the $p$-type sensors at a fluence of $5\times10^{14}~\textrm{n}_\textrm{eq}\textrm{cm}^{-2}$, 
%
while negligible differences in charge collection between the two sensor types were 
observed for thinner sensors 
at their 
respective expected maximum fluences.
%
The following results 
complete the study introduced in \cite{Peltola2017} 
and 
address the processes leading to the observations described above. 
%
%
The study involved 30 neutron irradiated test diode samples diced from 
8-inch (shallow diffused-FZ and epitaxial) 
and 
6-inch (standard and deep diffused-FZ) wafers. 

The experimental results 
were 
reproduced with 
Technology Computer-Aided Design (TCAD) simulations that allow reconstruction of 
the 
electric field distributions as well as the charge and current response of the 
irradiated test diodes within the validated fluence range of the 
defect model for neutrons \cite{Eber2013,Peltola2015r}. 
%
In the light of an earlier charge collection study at extreme fluences \cite{Kramberger2013}, the investigation of 200- 
and 300-$\upmu\textrm{m}$-thick $n$-on-$p$ sensors is extended beyond their respective maximum lifetime fluences up to about $1\times10^{16}~\textrm{n}_\textrm{eq}\textrm{cm}^{-2}$. 
The CCE evolution 
of the two sensor thicknesses is presented.

The 
radiation hardness 
results presented here cover only the bulk properties of the silicon sensor. Investigations of 
surface properties for the functionality of a multi-channel detector $-$such as of accumulation of Si/SiO$_2$ interface charge density with fluence, charge sharing, 
and inter-electrode capacitances and resistances$-$ are 
presently ongoing. 

The paper is arranged in following order: Section~\ref{SF} introduces the samples and target fluences involved in the neutron irradiation campaign. Measurement and simulation setups are presented in Section~\ref{Msetup}. In Section~\ref{Results}, characterization results of irradiated pad sensors are presented by first extracting effective fluences ($\Phi_\textrm{eff}$) from the leakage current volume density. Next, measured and simulated charge collection results are presented, followed by full depletion voltages ($V_\textrm{fd}$) extracted from the capacitance-voltage data. Finally, the results are discussed and summarised in Sections~\ref{Discussion} and~\ref{Conclusions}.

\section{Neutron irradiation campaign}
\label{SF}
\subsection{Samples and target fluences}
\label{StF}
The irradiation campaigns were carried out in reactors in 
two 
facilities, Rhode Island Nuclear Science Center\footnote{http://www.rinsc.ri.gov/} (RINSC) and UC Davis McClellan Nuclear Research Center\footnote{https://mnrc.ucdavis.edu/} (MNRC). 
The two independent irradiation runs, 16 samples at RINSC and 14 samples at MNRC, were included in the study in order to cross-check 
dosimetries of the facilities as well as 
the method 
by which the 
effective fluences are extracted from the leakage current ($I_{\textrm{leak}}$) of the 
samples after the irradiation. 

\begin{figure*}
     \centering
     \subfloat[]{\includegraphics[width=.74\textwidth]{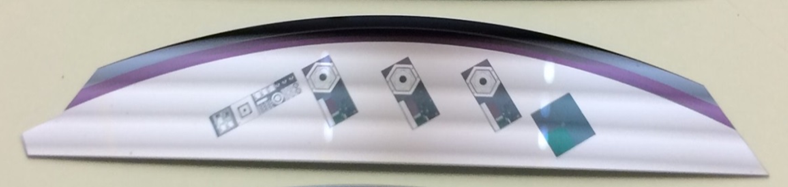}\label{8inSample}}\hspace{1mm}%
     \subfloat[]{\includegraphics[width=.2\textwidth]{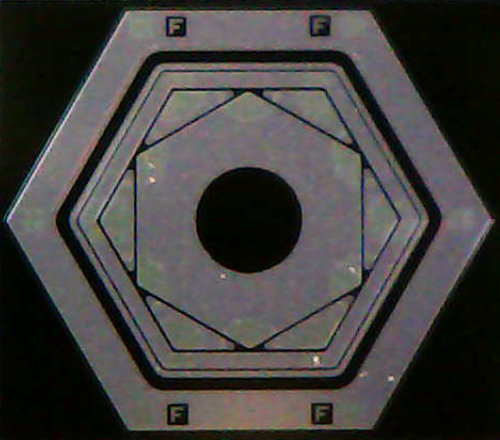}\label{TestDiode}}
    \caption{\small (a) Half moon sample with three test diodes and other test structures diced from an 8-inch Si wafer. (b) Close-up of the hexagonal 
test diode on the half moon sample. 
Corner to corner, the outer ring measures 4.5 mm. 
} 
\label{fig:fig14}
\end{figure*}

\begin{figure*}
\centering
\subfloat[]{\includegraphics[width=.47\textwidth]{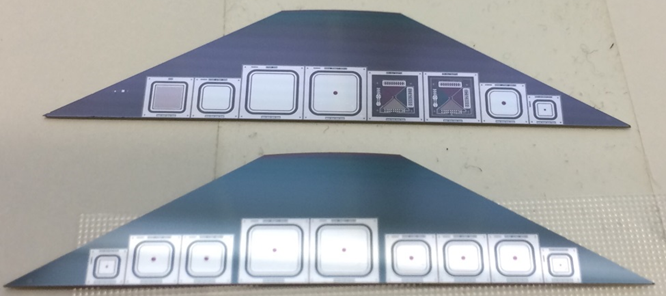}\label{DDsamples}}\hspace{1mm}%
\subfloat[]{\includegraphics[width=.495\textwidth]{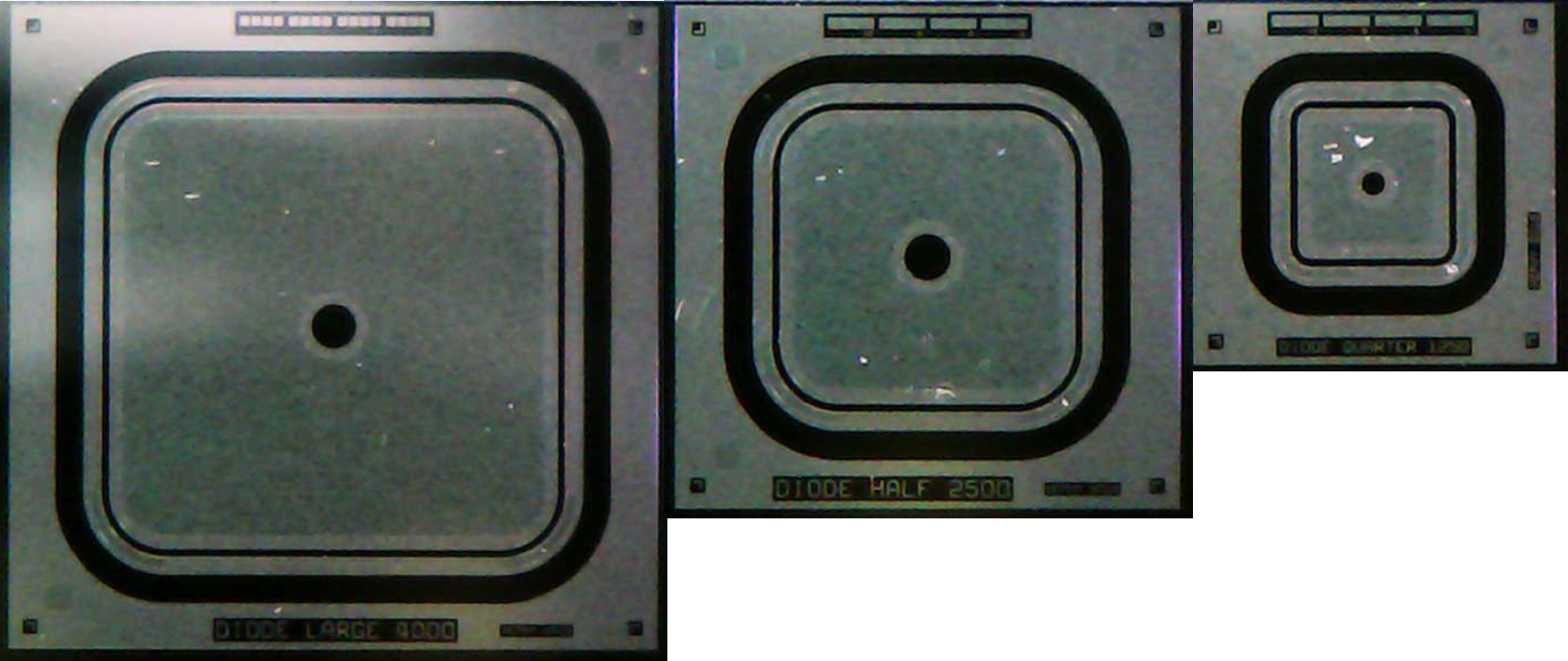}\label{gds}}%
\caption{\small (a) Half moon samples with test structures diced from a 6-inch Si wafer. Top: three test diodes, MOS- and other test structures. Bottom: nine test diodes. (b) Close-up of the three square diode sizes ($4.0\times4.0~\textrm{mm}^2$, $2.5\times2.5~\textrm{mm}^2$, $1.25\times1.25~\textrm{mm}^2$) with laser entrance windows in the center. 
}
\label{Samples}
\end{figure*}

Samples in Figures~\ref{fig:fig14} and~\ref{Samples} were diced from 8-inch and 6-inch Hamamatsu Photonics K.K. (HPK) 
sensors. 
Since the deep diffusion process used by HPK for reducing the active thickness to 300, 200, and 120 $\upmu\textrm{m}$, while maintaining
the physical thickness to 320 $\upmu\textrm{m}$, 
was only applied to the 6-inch wafers, the 8-inch 200- and 300-$\upmu\textrm{m}$-thick sensors have
the same physical and active thickness. 
Three distinct diffusion processes were applied to produce the heavily-doped backplane blocking contact: in 6-inch wafers standard-diffusion (`std-FZ') to produce 300-$\upmu\textrm{m}$ active thickness sensors and deep-diffusion (`dd-FZ') for 200- and 120-$\upmu\textrm{m}$ active thickness, while the 8-inch wafers applied shallow-diffusion (`shd-FZ') for 300- and 200-$\upmu\textrm{m}$-thick sensors. 
The only 8-inch wafer sensor in the study with a nominal active thickness of 120 $\upmu\textrm{m}$ 
was produced by growing epitaxial
silicon on a lower resistivity substrate, resulting in a total physical thickness of 300 $\upmu\textrm{m}$ \cite{Phase2}. While the substrate material of both 6- and 8-inch wafers is float-zone Si, the 
8-inch wafers have roughly an-order-of-magnitude lower oxygen concentration in the bulk relative to the 6-inch wafers. All the samples in 
the irradiation 
campaign are listed in Table~\ref{table_fs}. The samples are identified in Section~\ref{Results} by e.g. 
`shd-FZ$_{-}$LO$_{-}$300P' that refers to a 300-$\upmu\textrm{m}$-active thickness float-zone $n$-on-$p$ type sensor with shallow diffused backplane implant and low oxygen bulk concentration (all varieties are presented in Table~\ref{table_fs}). 

The 16 samples irradiated at the RINSC reactor 
were exposed at all 
six 
fluence levels shown in Table~\ref{table_fs}. 
%
The facility's dosimetry is based on the activation data of ultrapure foils of several different elements. Neutron flux spectrum of the reactor is extracted by using the known cross section of these elements to 
be activated by neutrons as a function of neutron energy. 
This measured neutron flux spectrum is then convoluted with the damage function for silicon \cite{Lazo1987} and integrated to obtain the 1-MeV neutron equivalent flux of $\phi=(4.0\pm0.8)\times10^{11}~\textrm{cm}^{-2}\textrm{s}^{-1}$.
%
For the six irradiation times ranging from 5.8 min to 383.2 min, this method 
resulted in about 20\% uncertainty in the dosimetry determined fluences.

The 14 samples irradiated at the MNRC reactor were exposed to the four highest 
fluences listed in Table~\ref{table_fs}. 
The first fluence was measured to be $7.37\times10^{14}$ 
n$_\textrm{eq}$cm$^\textrm{-2}$ using the fast sulfur activation method. The errors associated with this method are 
counting uncertainty in measuring the activation of 
$^{32}$P ($<0.5\%$), and the small non-uniformity in the flux along the length of the samples which 
contributes at most 5\%.
%
Since higher fluences would have made the sulfur samples too radioactive, 
the other three fluences 
were estimated based on the dosimetry results from the first irradiation. 
The reactor power levels were increased linearly to 
reach the higher fluences, resulting in 
a maximum 
uncertainty of 5\% 
associated with the sample position, 
with additional 5\% 
uncertainty associated with the normal power fluctuations in the reactor. 
When 
the 
uncertainties were assumed to be independent, the maximum dosimetry 
uncertainty expected for the three higher power irradiations was at most 15\%.

The dosimetry-determined fluences from both irradiation facilities, along with the target fluences and the leakage current extracted effective fluences, are 
further discussed in Section~\ref{secLCV}.

After irradiation, 
the samples were shipped in thermally isolated containers 
packed with 
cold gel to 
Texas Tech University (TTU) 
to avoid 
annealing of the radiation induced defects, 
and were then kept at 
-40 $^{\circ}$C 
at all times between the measurements.
%
\begin{table*}[!t]
\centering
\caption{Target fluences 
and irradiated samples. 
Bulk material is indicated as `shd-', `std-' and `dd-FZ' that correspond to shallow-, standard- and deep-diffused float zone, respectively, and `epi' for epitaxial. 
High and low oxygen bulk concentrations are indicated as `HO' and `LO', respectively. 
The 
two values in each entry in columns 3--8 indicate the number of samples irradiated at RINSC and MNRC reactors, respectively. The `wafer size' 
refers to the size of the full wafers from which the test samples were diced off.}
\label{table_fs}
\begin{tabular}{|c|c|c|c|c|c|c|c|}
\hline
\multirow{2}{*}{\textbf{Wafer}} & \multirow{2}{*}{\textbf{Sensor type \&}} & \multicolumn{6}{c|}{\textbf{Target fluence} [n$_\textrm{eq}$cm$^\textrm{-2}$]}\\
\cline{3-8}
\multicolumn{1}{|c|}{\textbf{size}} & \multicolumn{1}{c|}{\textbf{thickness}} & \multicolumn{1}{c|}{{\small $1.5\times10^{14}$}} & \multicolumn{1}{c|}{{\small $5.0\times10^{14}$}} & \multicolumn{1}{c|}{{\small $7.5\times10^{14}$}} & \multicolumn{1}{c|}{{\small $1.5\times10^{15}$}} & \multicolumn{1}{c|}{{\small $3.8\times10^{15}$}} & \multicolumn{1}{c|}{{\small $1.0\times10^{16}$}}\\
\hline
\multirow{3}{*}{{\large 8''}} & {\small shd-FZ$_{-}$LO$_{-}$300P} & {\small 1 + 0} & {\small 1 + 0} & {\small 2 + 2} & & & {\small 0 + 1}\\
\cline{2-8}
& {\small shd-FZ$_{-}$LO$_{-}$200P} & & & & {\small 1 + 0} & {\small 1 + 1} & {\small 0 + 1}\\
\cline{2-8}
& {\small epi$_{-}$LO$_{-}$120P} & & & & & {\small 0 + 1} & \\
\hline
\multirow{5}{*}{{\large 6''}} & {\small std-FZ$_{-}$HO$_{-}$300P} & & & & {\small 0 + 1} & {\small 0 + 1} & {\small 0 + 1}\\
\cline{2-8}
& {\small dd-FZ$_{-}$HO$_{-}$120P} & & & & {\small 1 + 0} & {\small 1 + 0} & {\small 1 + 2}\\
\cline{2-8}
& {\small std-FZ$_{-}$HO$_{-}$300N} & {\small 1 + 0} & {\small 1 + 0} & {\small 1 + 1} & & & \\
\cline{2-8}
& {\small dd-FZ$_{-}$HO$_{-}$200N} & & & & {\small 1 + 0} & & \\
\cline{2-8}
& {\small dd-FZ$_{-}$HO$_{-}$120N} & & & & & {\small 2 + 0} & {\small 1 + 2}\\
\hline
\end{tabular}
\end{table*}
\section{Measurement and simulation setups}
\label{Msetup}
\subsection{CV/IV-probestation}
\label{CV/IVstation}
%
For the capacitance-voltage/current-voltage ($CV/IV$) characterization, a 
custom probe station has been constructed at 
TTU that provides cooling to about -10 $^{\circ}$C in a dry environment and bias voltages up to 2.0 kV for the measurements of heavily-irradiated Si sensors. 
%
The bias voltage 
is supplied 
by a Keithley 2410 SourceMeter (SMU) up to 1.1 kV (or two back-to-back connected SMUs up to 2.0 kV) while the leakage current and capacitance are read out by a Keithley 6485 Picoammeter and a HP4274A LCR-meter, respectively. The 
DC-separation up to 2.0 kV of the LCR-meter's high- and low-terminals 
is 
accomplished by 1-$\upmu\textrm{F}$ capacitors. The remote control and 
data acquisition functions are carried out with LabVIEW\texttrademark-based software.

The test diode 
is connected to the measurement circuit by a vacuum chuck from its backplane, which also provides fixed position, and by probe needles on the segmented front surface for the diode region and guard-ring, respectively. 
A Peltier cooler 
below the stainless-steel chuck 
provides 
cooling while the heat 
from the Peltier is 
removed by a closed circuit liquid cooling system. 
The sample temperature and the humidity inside the cooling box are monitored by a thermocouple and a Vaisala DSS70A dew point meter, respectively. 

Since full depletion voltage ($V_\textrm{fd}$) has negligible sensitivity to measurement temperature in the studied $T$-range, the $CV$-measurements were carried out 
below +5 $^{\circ}$C to suppress leakage current to $\leq$ 1 mA during $V$-ramping for low power dissipation. Due to high sensitivity of leakage current to the measurement temperature, 
the temperature during $IV$-measurements was fixed as closely as possible to +5 $^{\circ}$C.
%
%
\subsection{TCT-setup}
The infrared-Transient Current Technique (IR-TCT) setup constructed at TTU for the measurement campaign is based on a Particulars commercial system\footnote{http://www.particulars.si} and modified to accommodate the investigated samples from 8-inch and 6-inch wafers 
at -30 $^{\circ}$C. 
%
The 
setup includes an 
IR-laser pulse generator of 1.06 $\upmu\textrm{m}$ wavelength 
that is connected by an optical fiber to a beam expander, which was placed about
85 mm above the center of the investigated diode that was mounted on an $xy$-translation stage. The diode was biased at the
front side with a Keithley 2410 SMU 
and read out through a 
Bias-T with a 
53 dB amplifier 
and a 2.5 GHz Tektronix DPO 7254 oscilloscope with 10 GS/s and 50 $\Omega$ termination. 
The remote control and DAQ of the circuit components are provided by LabVIEW\texttrademark-based software. 

The IR-TCT setup 
produces well-defined and stable transient signals that can be closely reproduced by simulation, as shown in Figure~\ref{fig:fig13} (simulation parameters are introduced in Section~\ref{SimSetup}). The IR-laser 
mimics a 
MIP by penetrating the entire thickness of the silicon sensors, thus 
providing straightforward means for a 
CCE investigation. 
A wire-bond connection minimizing the signal path between the sensor 
and the SMA cable 
was found to be essential in minimizing signal undershoot and reflections. 

The setup utilizes four Peltier coolers 
to achieve -30 $^{\circ}$C 
sample temperature. 
As in the $CV/IV$-setup, to avoid any condensation, 
N$_2$ flow was supplied to the sample during cooling. 
The temperature and humidity monitoring are accomplished by a four-wire resistance temperature detector (RTD) circuit and a Vaisala DSS70A dew point meter, respectively, thus ensuring that all measurements were carried out above dew point temperature. 
\subsection{Simulation setup and parameters}
\label{SimSetup}
%
%
All simulations 
in this paper were carried out using the Synopsys Sentaurus\texttrademark\footnote{http://www.synopsys.com} finite-element 
TCAD software framework. 
For the simulation 
of the evolution of electrical properties and charge collection with fluence, 
the two dimensional structures presented in Figure~\ref{fig:8b} were applied. The sensor width 
was set to wide enough to fully contain the generated charge carrier clouds within the structure during the transient simulation, when the laser illumination was 
performed at the center of the front surface ($x=100$, $y=0$ in Figure~\ref{D5005}). The simulation was focused at the center part of the test diodes. Due to the large surface areas of the real diodes, 
any contribution from the edges to the local electric fields at the center were considered negligible, and no edge regions were included in the modeled structures.  

The parameters extracted from $CV/IV$- and TCT-measurements before irradiation (bulk doping, active thickness, charge carrier trapping times, and backplane doping profiles for deep-diffused diodes) were used as an input to reproduce the devices as closely as possible by simulation. 
Doping profiles of all modeled diodes are presented in Figure~\ref{D_120P}. Displayed in Figure~\ref{D_120P}, the thermal drive-in of dopants up to 200 $\upmu\textrm{m}$ from the backplane in deep-diffused sensors results in significantly less abrupt doping profiles to shallow-diffused and epitaxial sensors. 

All modeled sensors were DC-coupled (with no oxide layer between the collecting electrode and Si bulk) as the real 
devices. 
High potential 
was provided from the backplane contact, while the front surface 
electrode at zero potential 
was used 
for charge collection 
in transient simulations. 

As in the experimental setup, an IR-laser of 
1.06 $\upmu\textrm{m}$ wavelength and sufficient penetration depth to model a MIP 
was applied to generate charge carriers in the detector. The laser spot diameter was set to 11 $\upmu\textrm{m}$ 
and the pulse length to 0.45 ns to match the values used in the IR-TCT setup. 
The laser intensity was tuned to produce MIP equivalent collected charge (22 k$e$ in 300 $\upmu\textrm{m}$ of silicon), while generating low enough \textit{e--h} pair density (about 1100 $\upmu\textrm{m}^{-2}$) to avoid any space charge density current effects that might modify sensor properties during the transient simulation. The bias-T circuit was modeled by including a high-resistance bias circuit and a charge-collection circuit with 50 $\Omega$ termination to the DC- and AC-parts of the simulation, respectively.

\begin{figure*}
     \centering
     \subfloat[ 
     ]{\includegraphics[width=.57\textwidth]{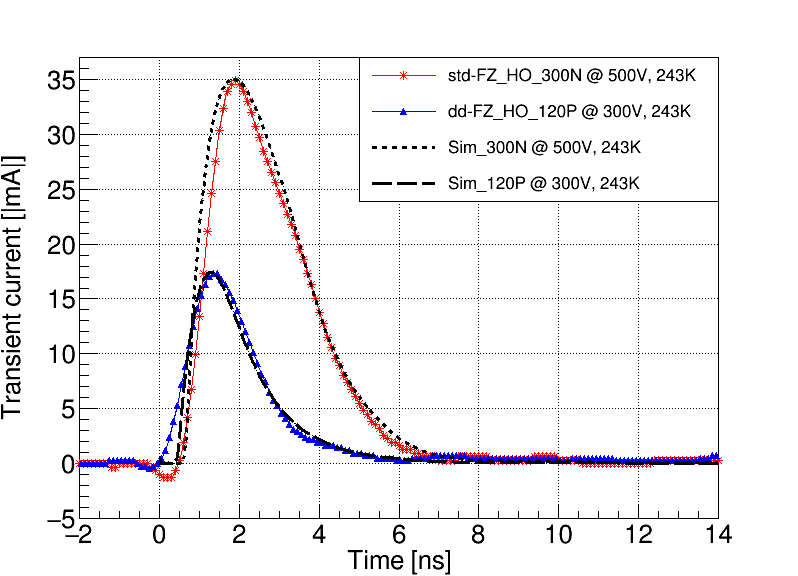}\label{IR_TCT}}
     \subfloat[ 
     ]{\includegraphics[width=.57\textwidth]{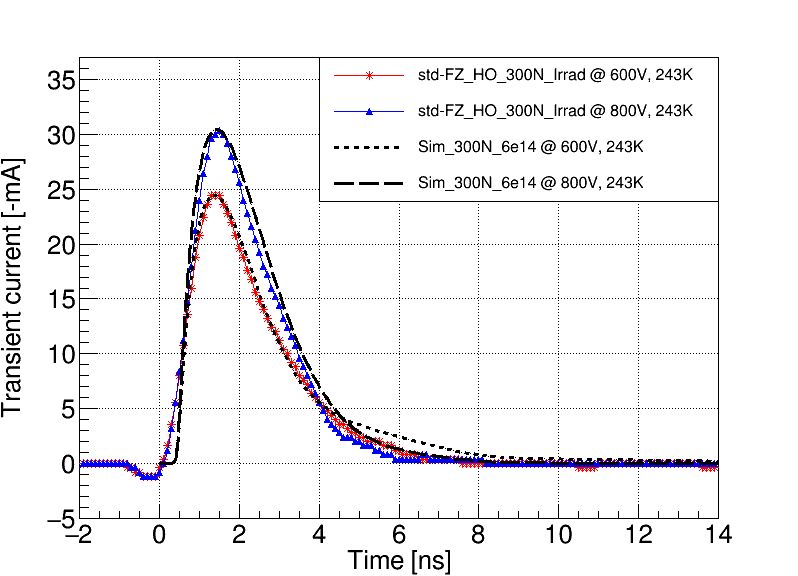}\label{TCT_Irrad}}\\
    \caption{\small Examples of measured and simulated transient signals for (a) pre-irradiated 300N and 120P sensors, and (b) 300N sensor after neutron irradiation to the fluence of $(6.1\pm0.5)\times10^{14}~\textrm{n}_\textrm{eq}\textrm{cm}^{-2}$. In (b) the simulation applied neutron defect model in Table~\ref{tabNM} with a fluence of $6.0\times10^{14}~\textrm{n}_\textrm{eq}\textrm{cm}^{-2}$. 
Both the measurements and the simulations were carried out with a 11 $\upmu\textrm{m}$ wide IR-laser spot, matching bias voltages and at -30 $^{\circ}$C. The sensor parameters used in the simulation were extracted from $CV/IV$-measurements.} 
\label{fig:fig13}
\end{figure*}

\begin{figure*}
     \centering
     \subfloat[]{\includegraphics[width=.39\textwidth]{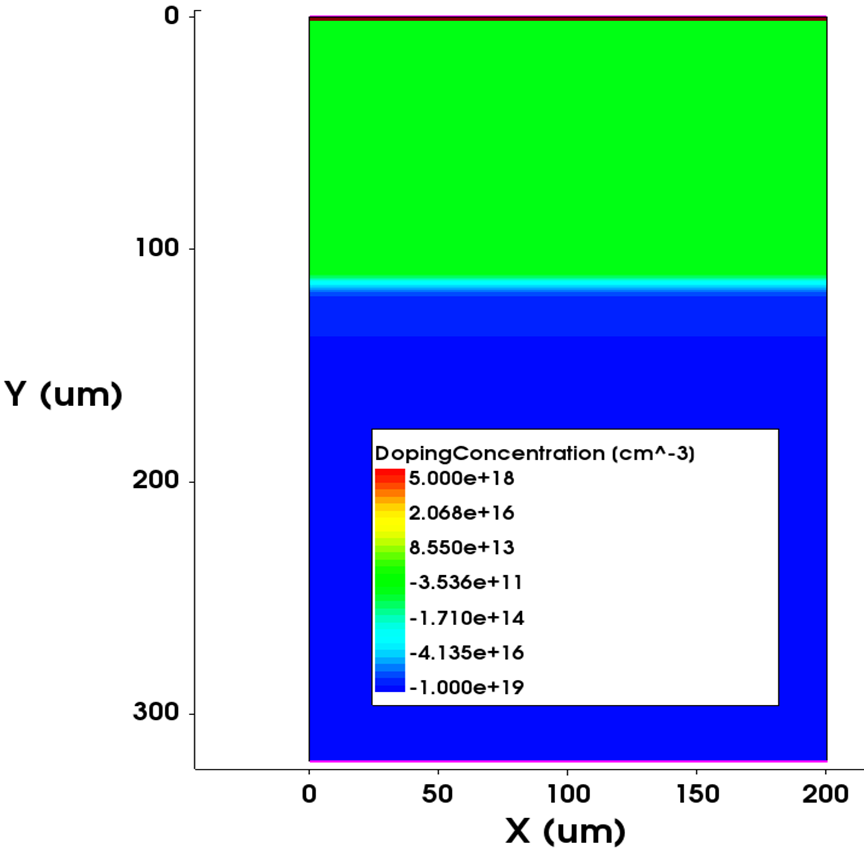}\label{D5005}}\hspace{1mm}%
     \subfloat[]{\includegraphics[width=.555\textwidth]{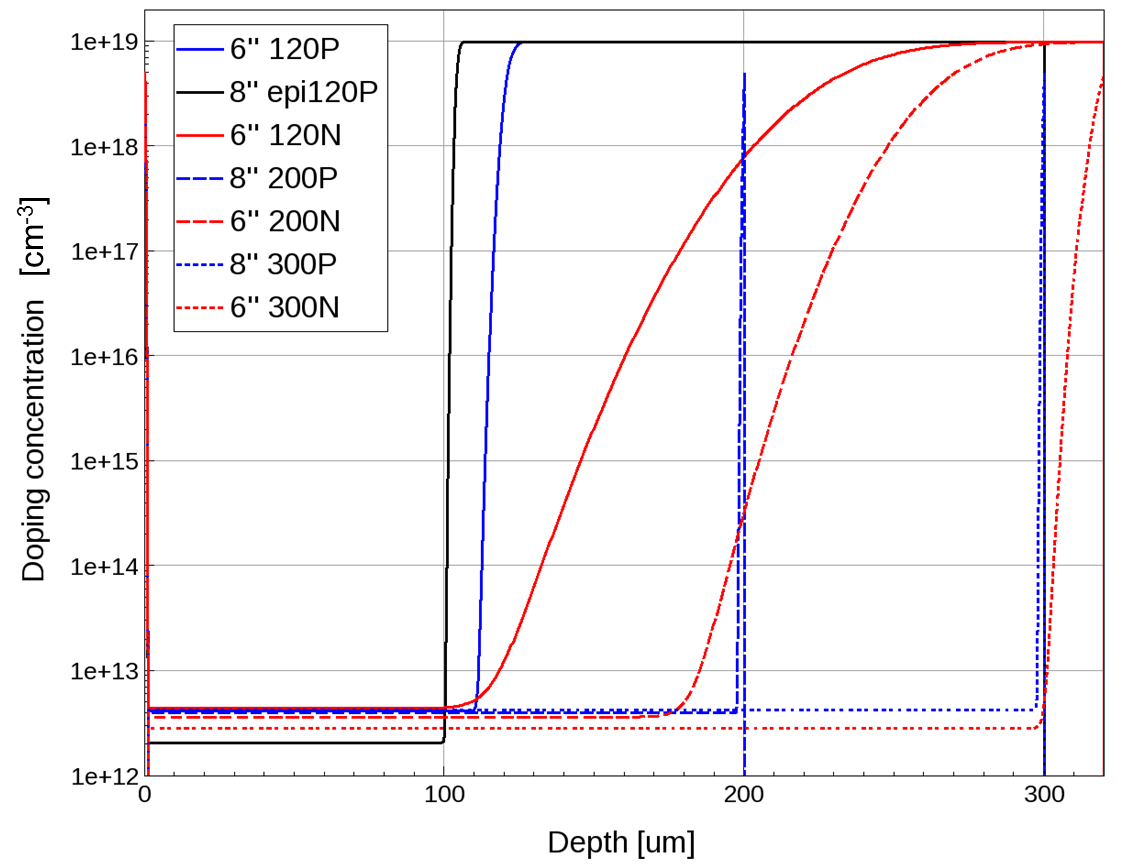}\label{D_120P}}\\
    \caption{\small 
Modeled sensor structures and doping profiles before irradiation. 
(a) A two dimensional $n$-on-$p$ sensor structure with 200 $\upmu\textrm{m}$ width, 320 $\upmu\textrm{m}$ physical thickness, 
and deep-diffusion thinned active thickness of 111 $\upmu\textrm{m}$. (b) 
Doping profiles of all simulated sensor structures. The active regions have doping levels of a few times $10^{12}~\textrm{cm}^{-3}$. 
} 
\label{fig:8b}
\end{figure*}

In heavily-irradiated silicon sensors 
(between $10^{14}$ to $10^{16}~\textrm{n}_\textrm{eq}\textrm{cm}^{-2}$), carrier trapping 
is the main factor reducing the collected charge. To model this in simulation, a neutron defect model \cite{Eber2013}, validated from $1\times10^{14}~\textrm{n}_\textrm{eq}\textrm{cm}^{-2}$ up to $1\times10^{15}~\textrm{n}_\textrm{eq}\textrm{cm}^{-2}$ and presented in Table~\ref{tabNM}, was applied. Figure~\ref{TCT_Irrad} displays examples of closely reproduced transient signals by the model for a neutron irradiated sensor. The neutron defect model was chosen over other models for higher fluences (e.g. \cite{Passeri2016}) 
because, at the moment, it is the only model that has been shown to closely reproduce the experimentally observed evolution with fluence of leakage current, full depletion voltage, and double-peak electric field distribution in the Si bulk, as well as charge collection efficiency. Although the model still reproduces expected leakage current well beyond the 
$1\times10^{15}~\textrm{n}_\textrm{eq}\textrm{cm}^{-2}$ limit over which it has been validated, the effective bulk doping ($N_\textrm{eff}$) increase with fluence becomes exaggerated, leading to high enough electric fields to generate \textit{e--h} avalanche that in turn leads to CCE values not in line with experimental observations. 
%
%
All charge collection simulations were carried out at matching temperature ($T=243~\textrm{K}$) to the measurements.

%
%
\begin{table*}[!t]
\centering
\caption{The parameters of the neutron defect model for Synopsys Sentaurus\texttrademark TCAD \cite{Eber2013}. $E_{\textrm{\small c,v}}$ are the conduction band and valence band energies, $\sigma$$_{\textrm{\small e,h}}$ are the electron and hole trapping cross sections, and $\Phi$ is the 1-MeV neutron equivalent fluence.} 
\label{tabNM}
\begin{tabular}{|c|c|c|c|}
    \hline
    {\bf Defect type} & {\bf Level} \textrm{[eV]} & {\bf $\sigma_{\textrm{\small e,h}}$} \textrm{[cm$^{2}$]} & 
{\bf Concentration} \textrm{[cm$^{-3}$]}\\
    \hline
    Deep acceptor &  $E_{\textrm{\small c}}-0.525$ & $1.2\times10^{-14}$ & $1.55\times\Phi$\\
    Deep donor &  $E_{\textrm{\small v}}+0.48$ & $1.2\times10^{-14}$ & $1.395\times\Phi$\\
    \hline
\end{tabular}
\end{table*}
\section{Characterization after irradiation}
\label{Results}
%
\subsection{Leakage currents and effective fluences}
\label{secLCV}
The leakage current ($I_\textrm{leak}$) measurements of the irradiated samples were carried out at +5 $^{\circ}$C up to a voltage of 1.1 kV. 
The $I_\textrm{leak}$ of fully depleted diodes, monitored by $CV$-measurements and discussed further in Section~\ref{secVfd}, was then used to determine the leakage current volume densities. 
For the active volumes, the diode area ($A$) information shown in Figures~\ref{fig:fig14} and~\ref{Samples} was used in combination with active thicknesses ($d$) extracted from measured geometrical capacitances ($C_\textrm{geom}=\epsilon_{0}\epsilon_\textrm{s}A/d$, where $\epsilon_{0}$ and $\epsilon_\textrm{s}$ are the vacuum permittivity and the relative permittivity of silicon, respectively \cite{Sze1981}).

%
\begin{linenomath}
Since current-related damage rate $\alpha$ has been shown to be strictly proportional to the 1-MeV neutron equivalent fluence as well as being independent of the silicon material, it is defined by \cite{Mollphd1999} 
\begin{equation}\label{eq1}
\frac{\Delta{I_\textrm{leak}}}{V}=\alpha\Phi_\textrm{eff},
\end{equation}
where $\Delta{I_\textrm{leak}}$ is the change in the leakage current due to irradiation, $V$ is the active volume of either $p$- or $n$-type detector, and $\Phi_\textrm{eff}$ is the 1-MeV neutron equivalent effective fluence. 
By using Eq.~\ref{eq1}, the 
$\Phi_\textrm{eff}$ extracted from $\Delta{I_\textrm{leak}}/V$ was determined by using 
$\alpha(\textrm{293 K})=4.0\times10^{-17}~\textrm{A/cm}$, a value established in previous studies \cite{Mollphd1999,Lindstrom2003}. To scale the +5 $^{\circ}$C measured leakage currents to room temperature, a method described in \cite{Moll1999} was applied. 
Several diodes were measured, and the final $\Phi_\textrm{eff}$ was determined as the mean value for all the samples 
corresponding to a given target fluence at a respective 
irradiation facility, as shown in Table~\ref{table_fs}.  
\end{linenomath} 

The results, along with the fluences determined by the dosimetry of the two irradiation facilities, 
are presented in Table~\ref{table_fluence}. 
The $\alpha$-values extracted from the slope of measured leakage current volume density as a function of $\Phi_\textrm{dosi}$ yield for RINSC and MNRC $(3.2\pm1.1)\times10^{-17}~\textrm{A/cm}$ and $(3.9\pm0.9)\times10^{-17}~\textrm{A/cm}$, respectively, matching $\alpha(\textrm{293 K})$ within uncertainty. Alternatively, when $\Phi_\textrm{eff}$ is plotted as a function of $\Phi_\textrm{dosi}$, the resulting slopes for RINSC and MNRC are $0.8\pm0.3$ and $1.0\pm0.3$, respectively, displaying agreement within uncertainty between data sets. 

All the 
measurements in this and the following sections were 
taken after 
about 10 min at 60 $^{\circ}$C, which is 
identical to the approach 
described in a previous Si sensor radiation hardness study for HGCAL \cite{Curras2017}. Additionally, after all samples had been TCT- and $CV$-characterized,
five samples 
exposed to four different fluences were annealed 80 min at 
60 $^{\circ}$C and $IV$-measured again. The resulting 
$\alpha(\textrm{293 K})$-factors extracted before and after the 80 min annealing for the four fluence points were $(3.83\pm0.74)\times10^{-17}~\textrm{A/cm}$ and $(3.79\pm1.12)\times10^{-17}~\textrm{A/cm}$, respectively. Thus, the original $IV$-extracted values were considered to provide a fair estimate of $\Phi_\textrm{eff}$ and are applied in fluence evolution plots of CCE and $V_\textrm{fd}$ in the following sections.
\begin{table*}[!t]
\centering
\caption{Target fluences, dosimetry fluences ($\Phi_\textrm{dosi}$) provided by the irradiation facilities, and $IV$-extracted effective fluences ($\Phi_\textrm{eff}$). All fluences are given in 1-MeV neutron equivalent units. 
} 
\label{table_fluence}
\begin{tabular}{|c|c|c|c|}
\hline
\multirow{2}{*}{{\bf Facility}} & \multirow{2}{*}[0.5em]{{\bf Target $\Phi$}} & \multirow{2}{*}[0.5em]{{\bf $\Phi_\textrm{dosi}$}} & \multirow{2}{*}[0.5em]{{\bf $\Phi_\textrm{eff}$}}\\
& [n$_\textrm{eq}$cm$^\textrm{-2}$] & [n$_\textrm{eq}$cm$^\textrm{-2}$] & [n$_\textrm{eq}$cm$^\textrm{-2}$]\\
\hline
\multirow{6}{*}{{\large RINSC}} & $1.5\times10^{14}$ & $(1.5\pm0.3)\times10^{14}$ & $(1.05\pm0.05)\times10^{14}$\\
\cline{2-4}
& $5.0\times10^{14}$ & $(4.6\pm0.9)\times10^{14}$ & $(3.5\pm0.4)\times10^{14}$\\
\cline{2-4}
& $7.5\times10^{14}$ & $(7.1\pm1.6)\times10^{14}$ & $(5.4\pm0.4)\times10^{14}$\\
\cline{2-4}
& $1.5\times10^{15}$ & $(1.3\pm0.3)\times10^{15}$ & $(1.5\pm0.3)\times10^{15}$\\
\cline{2-4}
& $3.8\times10^{15}$ & $(3.4\pm0.7)\times10^{15}$ & $(2.35\pm0.19)\times10^{15}$\\
\cline{2-4}
& $1.0\times10^{16}$ & $(8.4\pm1.8)\times10^{15}$ & $(6.6\pm0.7)\times10^{15}$\\
\hline
\multirow{4}{*}{{\large MNRC}} & $7.5\times10^{14}$ & $(7.4\pm1.1)\times10^{14}$ & $(6.1\pm0.5)\times10^{14}$\\
\cline{2-4}
& $1.5\times10^{15}$ & $(1.47\pm0.22)\times10^{15}$ & $(1.3\pm0.3)\times10^{15}$\\
\cline{2-4}
& $3.8\times10^{15}$ & $(3.7\pm0.6)\times10^{15}$ & $(3.47\pm0.16)\times10^{15}$\\
\cline{2-4}
& $1.0\times10^{16}$ & $(9.8\pm1.5)\times10^{15}$ & $(9.3\pm1.1)\times10^{15}$\\
\hline
\end{tabular}
\end{table*}
%
%
%
\subsection{Charge collection}
\subsubsection{Extracting CCE from IR-TCT data}
To extract the 
CCE of the irradiated diodes, the IR-laser induced transient currents were first recorded at -30 $^{\circ}$C for both the irradiated 
and non-irradiated-reference samples. This was done to compensate for the observed temperature dependence of the transient-signal amplitudes. The reference and the corresponding irradiated sample 
were always diced from the same Si wafer to minimize any effect from possible processing differences between the wafers. 

%
Transient currents of the reference diodes were recorded for reverse bias voltages from 50 V above $V_\textrm{fd}$ up to 500 V. The laser spot was focused at the center of the laser window (see Figures~\ref{fig:fig14} and~\ref{Samples}) and the 
light intensity was set as low as possible, while not compromising the stability of the transient signal, to avoid any space charge density current effects 
as well as amplifier saturation. 
The collected charge ($Q_\textrm{coll}$) was then 
extracted by integrating the current signal over time.
%
The most probable value of $Q_\textrm{coll}$ in a fully depleted reference diode was computed 
as an average of collected charges recorded over all voltages. 

28 irradiated samples were included in the CCE study. The investigated voltages for the irradiated samples ranged from 400 V to 1 kV and the same analysis methods described above for reference diodes were applied.
The charge collection efficiency was then determined as a ratio CCE = $Q_\textrm{coll}$(irradiated)/$Q_\textrm{coll}$(reference). The RTD-monitored 
mean temperature 
throughout all measurements was $T=($-30.10$\pm$0.13$)$ $^{\circ}$C. 
The CCE results 
were also converted to the collected charge by considering a 
MIP-induced charge deposition of 73 $e/\upmu\textrm{m}$ in 300 $\upmu\textrm{m}$ silicon, and 75 $e/\upmu\textrm{m}$ in 120--200 $\upmu\textrm{m}$ silicon 
\cite{Bichsel1988,Patri2016}. Both are presented in the following section. 

%
\subsubsection{CCE results}
\label{CCEresults}
\textit{300-$\mu\textrm{m}$-thick sensors}

\begin{figure*}
     \centering
     \subfloat[ 
     ]{\includegraphics[width=.70\textwidth]{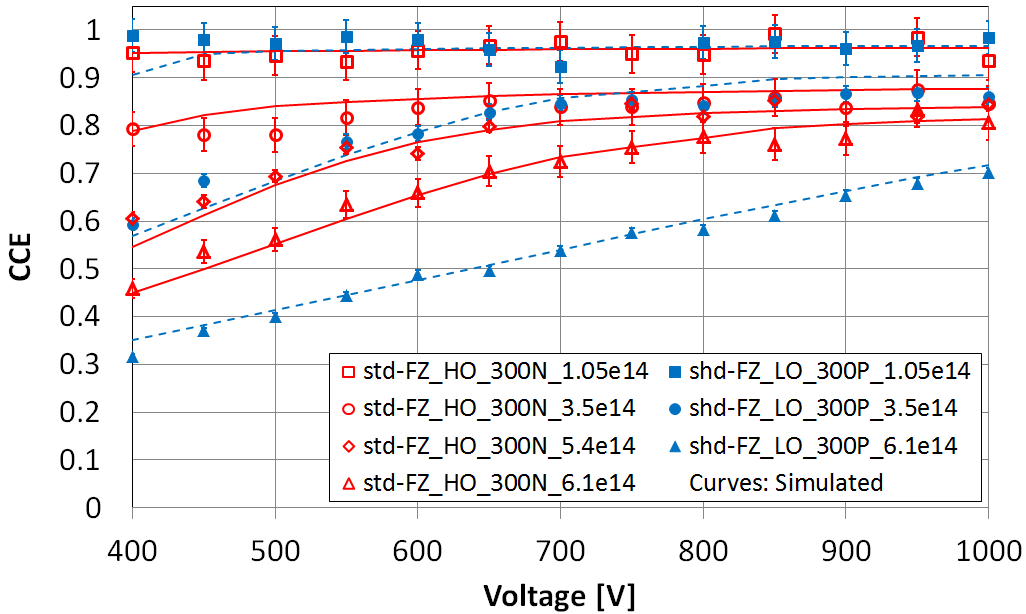}\label{CCE_V300}}\\
     \subfloat[ 
     ]{\includegraphics[width=.70\textwidth]{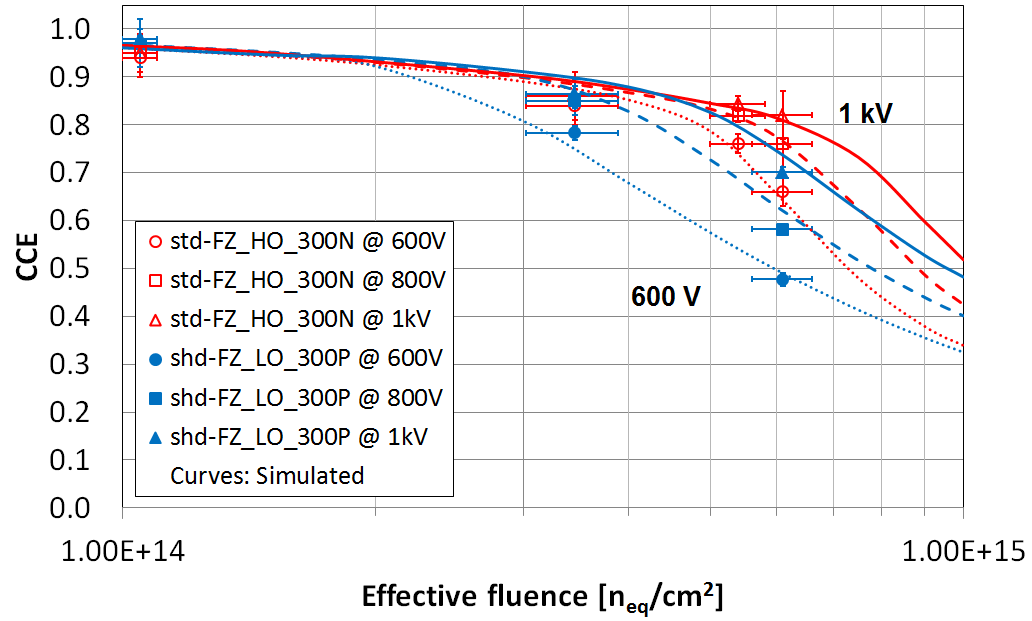}\label{CCE_F300}}\\
    \caption{\small 
Evolution of CCE 
at 
lowest-fluence region 
with voltage and fluence for 300-$\upmu\textrm{m}$ thick sensors at -30 $^{\circ}$C. 
Sensor identification is indicated in the legends, as well as effective fluences (in units of n$_\textrm{eq}$cm$^\textrm{-2}$) in panel (a). 
(a) Measured 
and simulated (
solid and dashed curves for 
$p$-on-$n$ and $n$-on-$p$, respectively, 
with $IV$-extracted effective fluences used as simulation input shown in Table~\ref{table_V300}) CCE($V$). 
(b) 
Measured 
and simulated (dotted for 600 V, dashed for 800 V and solid curves for 1 kV) 
CCE($\Phi_\textrm{eff}$). 
} 
\label{fig_CCE300}
\end{figure*}

In Figure~\ref{fig_CCE300}, 
the measured results for 300-$\upmu\textrm{m}$ thick diodes also include comparison with TCAD simulated CCE. 
All expected lifetime fluences for 300-$\upmu\textrm{m}$ thick sensors at HGCAL, from $1\times10^{14}~\textrm{n}_\textrm{eq}\textrm{cm}^{-2}$ to about $(5-6)\times10^{14}~\textrm{n}_\textrm{eq}\textrm{cm}^{-2}$ \cite{Phase2}, 
are within the validated range of the neutron defect model. 
The simulation closely reproduces 
both voltage (with input fluences within uncertainty of $\Phi_\textrm{eff}$, as shown in Table~\ref{table_V300}) and fluence dependence of the measured CCE. 
%

Figure~\ref{CCE_F300} displays significantly better CCE performance for 300N (300-$\upmu\textrm{m}$ thick $p$-on-$n$) sensors compared to 300P above fluence of about $4\times10^{14}~\textrm{n}_\textrm{eq}\textrm{cm}^{-2}$. 
This is due to space charge sign inversion (SCSI) of the $n$-type substrate that results in the $p$-on-$n$ sensor being fully depleted at lower voltages, despite the $pn$-junction has moved to the backside, than the $n$-on-$p$ in the fluence range 
of 300 $\upmu\textrm{m}$ thick sensors. 
%
At fluences around $6\times10^{14}~\textrm{n}_\textrm{eq}\textrm{cm}^{-2}$, the CCEs of both polarity sensors exhibit 10\% and $\sim$20\% gains from increasing the operating voltage from 600 V to 800 V and 1 kV, respectively.\\ \\
%
%
\begin{table*}[!t]
\centering
\caption{Sensor identification, $IV$-extracted effective fluences ($\Phi_\textrm{eff}$) 
and simulation input fluences ($\Phi_\textrm{{\tiny TCAD}}$) for the CCE results in Figure~\ref{CCE_V300}. Bulk material is indicated as `shd-' and `std-FZ' for shallow- and standard-diffused float zone, respectively. 
High (`HO') and low (`LO') oxygen bulk concentrations correspond to samples diced from 6- and 8-inch wafers, respectively. 
} 
\label{table_V300}
\begin{tabular}{|c|c|c|}
\hline
\multirow{2}{*}[0.5em]{{\bf Sensor thickness \&}} & \multirow{2}{*}[0.5em]{{\bf $\Phi_\textrm{eff}$}} 
& \multirow{2}{*}[0.5em]{{\bf $\Phi_\textrm{{\tiny TCAD}}$}}\\
{\bf type} & [n$_\textrm{eq}$cm$^\textrm{-2}$] & [n$_\textrm{eq}$cm$^\textrm{-2}$]\\
\hline
shd-FZ$_{-}$LO$_{-}$300P & $(1.05\pm0.05)\times10^{14}$ & $1.1\times10^{14}$\\
\hline
std-FZ$_{-}$HO$_{-}$300N & $(1.05\pm0.05)\times10^{14}$ & $1.1\times10^{14}$\\
\hline
shd-FZ$_{-}$LO$_{-}$300P & $(3.5\pm0.4)\times10^{14}$ & $3.2\times10^{14}$\\
\hline
std-FZ$_{-}$HO$_{-}$300N & $(3.5\pm0.4)\times10^{14}$ & $3.9\times10^{14}$\\
\hline
std-FZ$_{-}$HO$_{-}$300N & $(5.4\pm0.4)\times10^{14}$ & $5.2\times10^{14}$\\
\hline
std-FZ$_{-}$HO$_{-}$300N & $(6.1\pm0.5)\times10^{14}$ & $6.0\times10^{14}$\\
\hline
shd-FZ$_{-}$LO$_{-}$300P & $(6.1\pm0.5)\times10^{14}$ & $6.2\times10^{14}$\\
\hline
\end{tabular}
\end{table*}
%
\textit{200-$\mu\textrm{m}$-thick sensors}

\begin{figure*}
     \centering
     \subfloat[ 
     ]{\includegraphics[width=.70\textwidth]{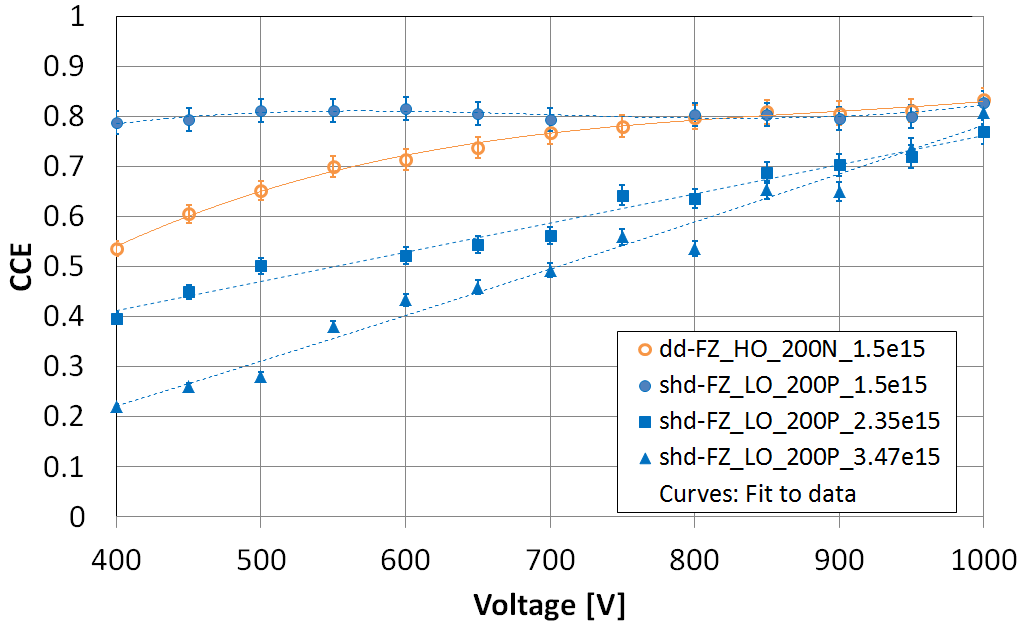}\label{CCE_V200}}\\
     \subfloat[ 
     ]{\includegraphics[width=.70\textwidth]{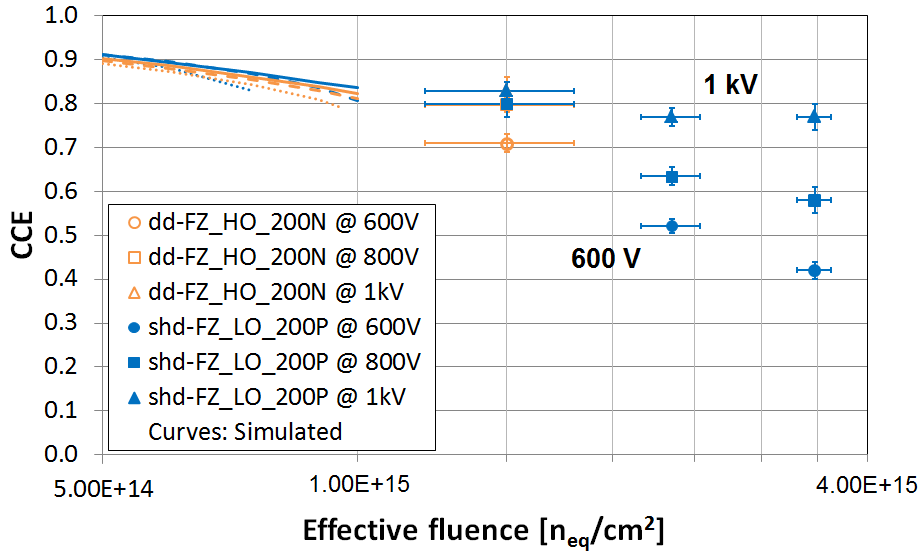}\label{CCE_F200}}
    \caption{\small 
CCE 
at 
intermediate-fluence region as a function of 
voltage and fluence for 200-$\upmu$m thick sensors at -30 $^{\circ}$C. 
Sensor identification is indicated in the legends, as well as effective fluences (in units of n$_\textrm{eq}$cm$^\textrm{-2}$) in panel (a). 
(a) Measured 
CCE($V$) 
with 
fits to data 
(polynomial and linear fits for the two lower-fluence and the two higher-fluence results, respectively) included. 
(b) 
Measured 
and simulated (dotted for 600 V, dashed for 800 V and solid curves for 1 kV) 
CCE($\Phi_\textrm{eff}$). } 
\label{fig_CCE200}
\end{figure*}

The results for 200-$\upmu\textrm{m}$-thick sensors (Figure~\ref{CCE_V200}) show voltage dependence of CCE for the three irradiated samples, 
which indicates that the sensors are not fully depleted below 800 V, as further discussed in Section~\ref{secVfd}. 
Because the lifetime fluences 
of 200-$\upmu\textrm{m}$-thick sensors 
are expected to be in the range of $(0.5-2.5)\times10^{15}~\textrm{n}_\textrm{eq}\textrm{cm}^{-2}$ \cite{Phase2}, the TCAD simulations in Figure~\ref{CCE_F200} are only applied to the lower part of the fluence range, up to $1\times10^{15}~\textrm{n}_\textrm{eq}\textrm{cm}^{-2}$. Since the lowest target fluence for the irradiated samples was $1.5\times10^{15}~\textrm{n}_\textrm{eq}\textrm{cm}^{-2}$, the simulations provide a complementary view of the CCE behavior throughout the expected fluence range when combined with 
data points of the 200P sensors. Since only one 200N sample was available for irradiation, 
the combined measured and simulated CCE(200N) results go up to $(1.5\pm0.3)\times10^{15}~\textrm{n}_\textrm{eq}\textrm{cm}^{-2}$. 

%
The CCE of 200P sensors 
does better by $\sim$11\% and 25\% 
at 
800 V and 1 kV, respectively, 
compared to at 600 V, close to the maximum expected lifetime fluence ($\Phi_\textrm{max}$) to which the 200-$\upmu\textrm{m}$ thick sensors will be exposed. 
Furthermore, at a fluence beyond $\Phi_\textrm{max}$, at $(3.47\pm0.16)\times10^{15}~\textrm{n}_\textrm{eq}\textrm{cm}^{-2}$, 
corresponding gains of 16\% and 35\% 
are observed. 

When extrapolated, the 
simulated CCE for the 200N sensor 
is consistent with the measured data points for 
the three voltages in Figure~\ref{CCE_F200}. 
%
By increasing the voltage from 600 V to 800 V and 1 kV, gains of $\sim$9\% and 12\% in CCE, respectively, are observed at the fluence of $(1.5\pm0.3)\times10^{15}~\textrm{n}_\textrm{eq}\textrm{cm}^{-2}$.\\ \\
%
\textit{120-$\mu\textrm{m}$-thick sensors}

\begin{figure*}
     \centering
     \subfloat[ 
     ]{\includegraphics[width=.70\textwidth]{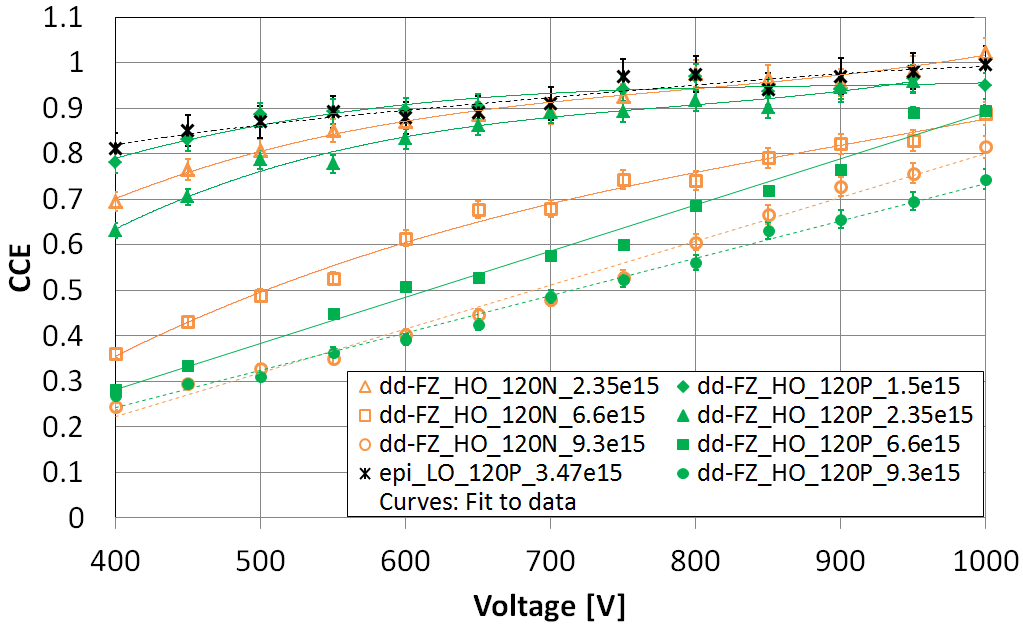}\label{CCE_V120}}\\
     \subfloat[ 
     ]{\includegraphics[width=.70\textwidth]{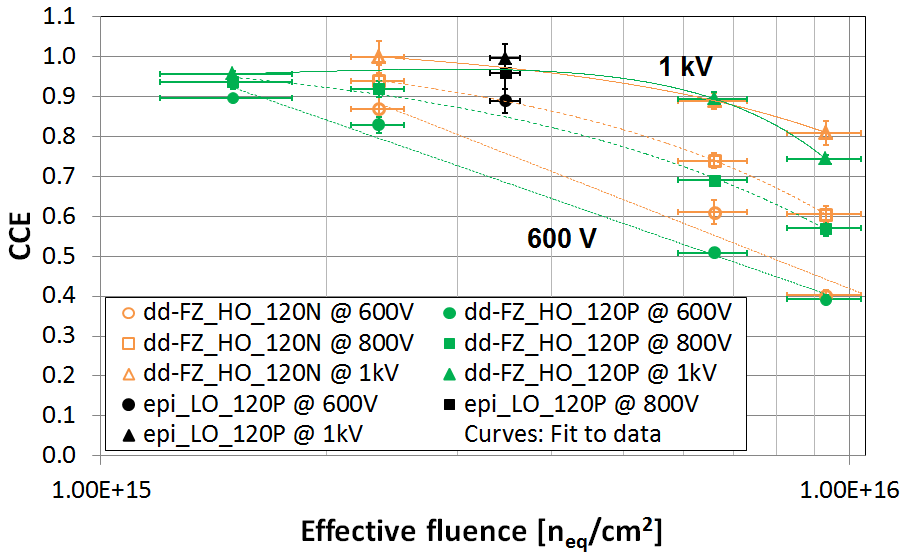}\label{CCE_F120}}
    \caption{\small 
CCE 
at 
highest-fluence region as a function of voltage and fluence for 120 $\upmu$m nominal thickness sensors at -30 $^{\circ}$C. 
Sensor identification is indicated in the legends, as well as effective fluences (in units of n$_\textrm{eq}$cm$^\textrm{-2}$) in panel (a). 
(a) Measured CCE($V$) 
with polynomial fits 
to data. 
(b) Measured 
CCE($\Phi_\textrm{eff}$) with corresponding 
logarithmic and polynomial fits for both dd-FZ sensor polarities. 
} 
\label{fig:fig12}
\end{figure*}

Since the 
120-$\upmu\textrm{m}$-thick sensors 
will have to operate in a neutron fluence range from about $2.5\times10^{15}~\textrm{n}_\textrm{eq}\textrm{cm}^{-2}$ 
to $1\times10^{16}~\textrm{n}_\textrm{eq}\textrm{cm}^{-2}$, 
which is well beyond the validated fluence limit of the TCAD neutron defect model, all the curves in Figure~\ref{fig:fig12} are fits to data points.

%
In Figure~\ref{CCE_V120},
the 
results from the 8-inch epitaxial 120P sample at the fluence of 
$(3.47\pm0.16)\times10^{15}~\textrm{n}_\textrm{eq}\textrm{cm}^{-2}$ display slightly higher CCE values 
than the two lower-fluence data sets from the dd-FZ samples.
The trend is similar to the results reported in \cite{Curras2017} for epitaxial 100P sensors and needs to be further investigated to be fully understood. Partial explanation is discussed in the end of Section~\ref{secVfd}. 

In Figure~\ref{CCE_F120}, 
the CCEs of both polarity sensors benefit $\sim$20\% and 40\% due to increased operating voltage from 600 V to 
800 V and 1 kV, respectively, at the highest fluence. 
Both polarities display similar CCE performance at the highest fluence. While it is not clear why 120N at the second highest fluence displays higher CCE values to 120P at lower voltages, the difference diminishes completely at higher voltages. \\ \\ 
%
\textit{Charge collection summary}

\begin{figure*} 
     \centering
     \subfloat[ 
     ]{\includegraphics[width=.70\textwidth]{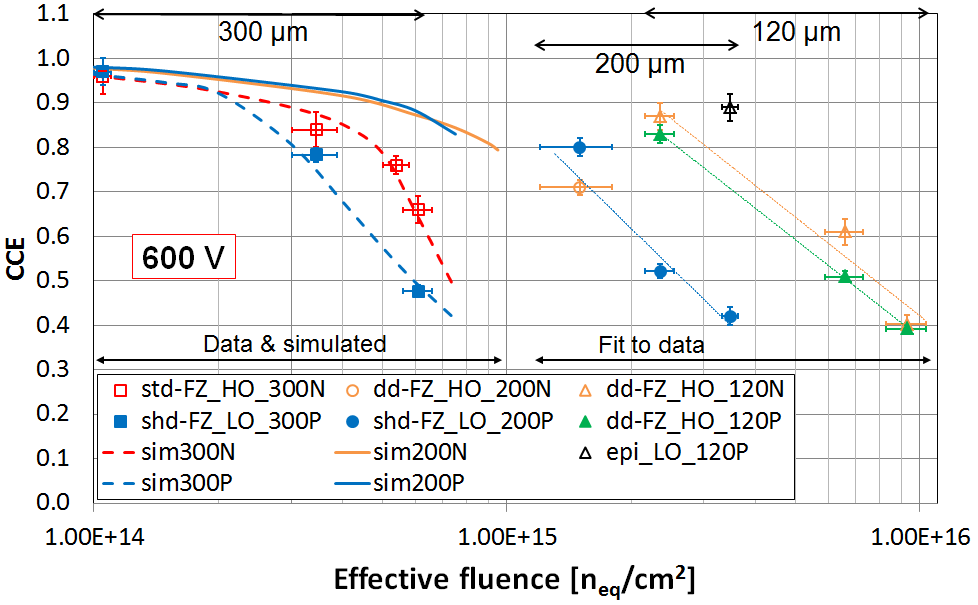}\label{CCE_123_600}}\\
     \subfloat[ 
     ]{\includegraphics[width=.70\textwidth]{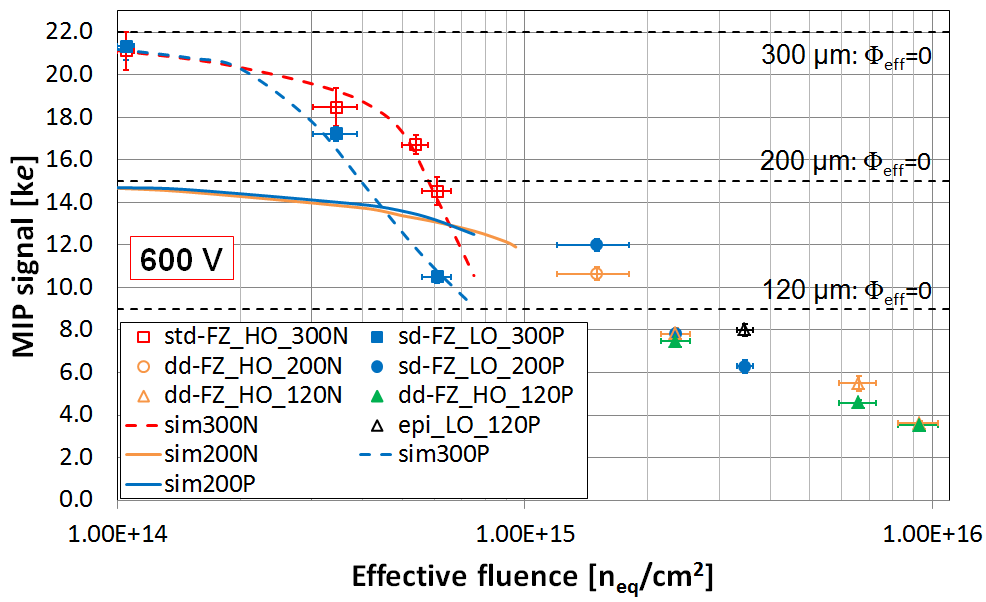}\label{ke_123_600}}
    \caption{\small 
CCE and MIP charge collection 
as a function of 
fluence 
for the three sensor thicknesses at 600 V and -30 $^{\circ}$C temperature, and for the 
full 
lifetime fluence range 
\cite{Phase2}. 
Sensor identification is indicated in the legends. 
(a) Measured and simulated 
CCE($\Phi_\textrm{eff}$) 
with the fluence range of the data points for each sensor thickness 
indicated. Also indicated, the curves below and above $1\times10^{15}~\textrm{n}_\textrm{eq}\textrm{cm}^{-2}$ are simulated results (thick solid and dashed) and logarithmic fits to data (thin dotted), respectively. 
(b)  Measured and simulated 
$Q_\textrm{coll}$($\Phi_\textrm{eff}$) from MIP charge deposition. The black dashed lines indicate the $Q_\textrm{coll}$ for each active thickness at $\textrm{CCE}=1$ \cite{Bichsel1988,Patri2016}. 
} 
\label{fig_CCE123}
\end{figure*}

\begin{figure*}
     \centering
     \subfloat[ 
     ]{\includegraphics[width=.70\textwidth]{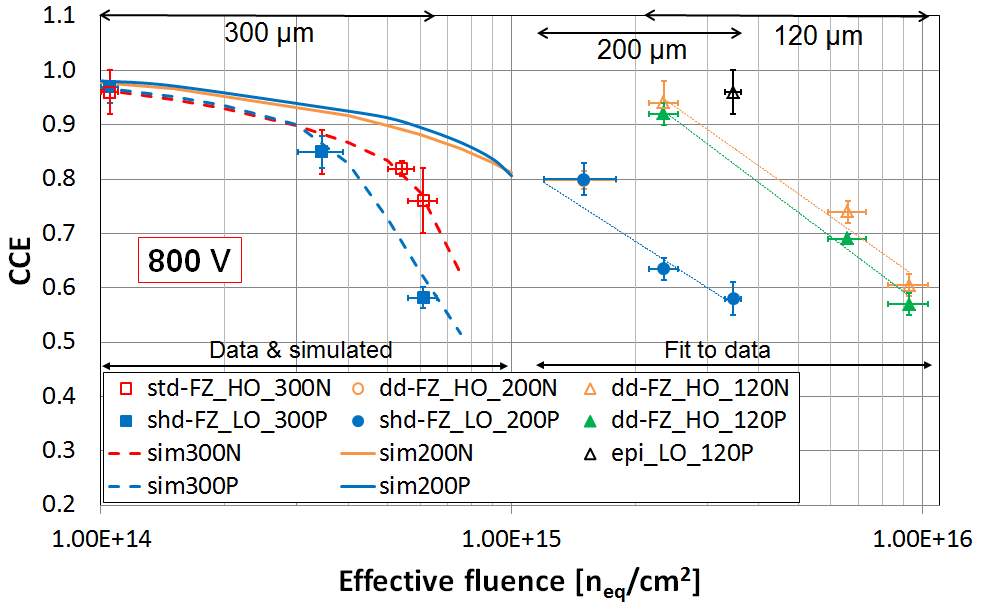}\label{CCE_123_800}}\\
     \subfloat[ 
     ]{\includegraphics[width=.70\textwidth]{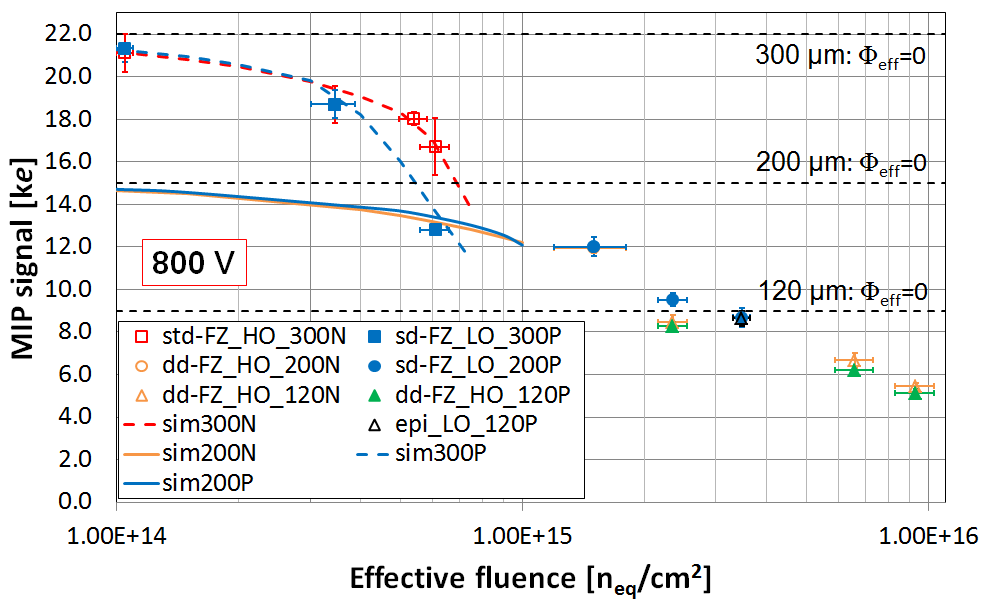}\label{ke_123_800}}
    \caption{\small 
Corresponding CCE and MIP charge collection 
as a function of 
fluence to Figure~\ref{fig_CCE123} 
at 800 V. 
Sensor identification is indicated in the legends. 
(a) Measured and simulated 
CCE($\Phi_\textrm{eff}$) 
with the fluence range of the data points for each sensor thickness 
indicated. Also indicated, the curves below and above $1\times10^{15}~\textrm{n}_\textrm{eq}\textrm{cm}^{-2}$ are simulated results (thick solid and dashed) and logarithmic fits to data (thin dotted), respectively. 
(b)  Measured and simulated 
$Q_\textrm{coll}$($\Phi_\textrm{eff}$) from MIP charge deposition with 
$Q_\textrm{coll}$ for each active thickness at $\textrm{CCE}=1$ indicated \cite{Bichsel1988,Patri2016}. 
} 
\label{fig_CCE123_800V}
\end{figure*}

\begin{figure*}
     \centering
     \subfloat[ 
     ]{\includegraphics[width=.70\textwidth]{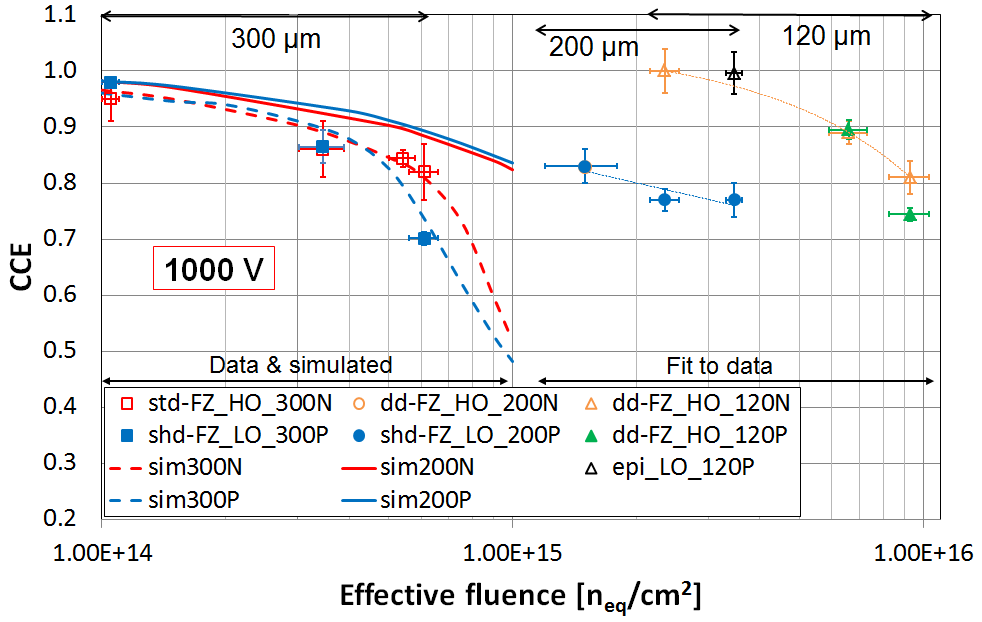}\label{CCE_123_1k}}\\
     \subfloat[ 
     ]{\includegraphics[width=.70\textwidth]{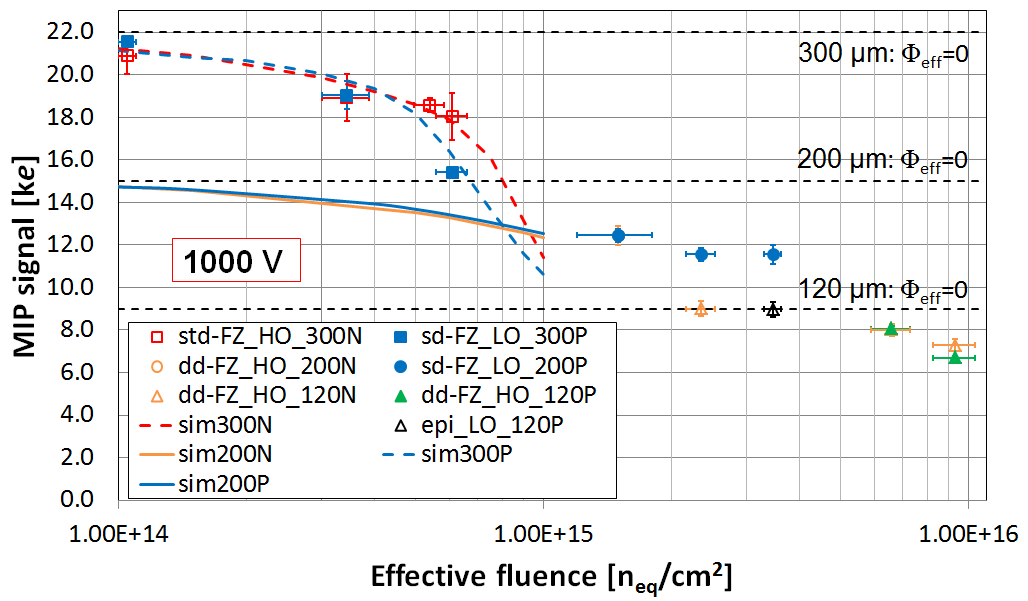}\label{ke_123_1k}}
    \caption{\small Corresponding CCE and MIP charge collection 
as a function of 
fluence to Figures~\ref{fig_CCE123} and~\ref{fig_CCE123_800V} 
at 1 kV. 
Sensor identification is indicated in the legends. 
(a) Measured and simulated 
CCE($\Phi_\textrm{eff}$) 
with the fluence range of the data points for each sensor thickness 
indicated. Also indicated, the curves below and above $1\times10^{15}~\textrm{n}_\textrm{eq}\textrm{cm}^{-2}$ are simulated results (thick solid and dashed) and logarithmic (200P) and polynomial (120N) fits to data (thin dotted), respectively. 
(b)  Measured and simulated 
$Q_\textrm{coll}$($\Phi_\textrm{eff}$) from MIP charge deposition with 
$Q_\textrm{coll}$ for each active thickness at $\textrm{CCE}=1$ indicated \cite{Bichsel1988,Patri2016}. 
} 
\label{fig_CCE123_1kV}
\end{figure*}

The measured CCE results for the three sensor thicknesses for the full expected lifetime fluence range are shown in Figures~\ref{CCE_123_600},~\ref{CCE_123_800}, and~\ref{CCE_123_1k}, while separating for the operating voltages of 600 V, 800 V, and 1 kV, respectively.
%
At fluences around 
$\Phi_\textrm{max}$ 
$\textrm{CCE}~{\geq}~60\%$ and over 70\% are observed at 800 V and 1 kV, respectively, for all measured sensors. 
When operated at 600 V, the same level of CCE performance at corresponding fluences is only seen for 300N sensors, 
as marked in Figure~\ref{CCE_123_600}. 
Close to lower lifetime fluence limits expected at the outer radii of each 
sensor thickness ($1\times10^{14}~\textrm{n}_\textrm{eq}\textrm{cm}^{-2}$ for 300 $\upmu\textrm{m}$, $5\times10^{14}~\textrm{n}_\textrm{eq}\textrm{cm}^{-2}$ for 200 $\upmu\textrm{m}$ and $2.5\times10^{15}~\textrm{n}_\textrm{eq}\textrm{cm}^{-2}$ for 120 $\upmu\textrm{m}$), operating at 600 V 
provides high CCE performance. 

The corresponding MIP-induced charge deposition results in Figures~\ref{ke_123_600},~\ref{ke_123_800}, and~\ref{ke_123_1k} indicate that the 120-$\upmu\textrm{m}$-thick sensors close to the fluence of $10^{16}~\textrm{n}_\textrm{eq}\textrm{cm}^{-2}$ are able to collect $\sim$3.6 at 600 V, 5.4 at 800 V, and 7.3 k$e$ at 1 kV of the 9 k$e$ generated in the silicon bulk. 
%
\subsubsection{Charge collection of 300P and 200P sensors at extreme fluences} 
\label{secCCE23}
\begin{figure*}
     \centering
     \subfloat[ 
     ]{\includegraphics[width=.49\textwidth]{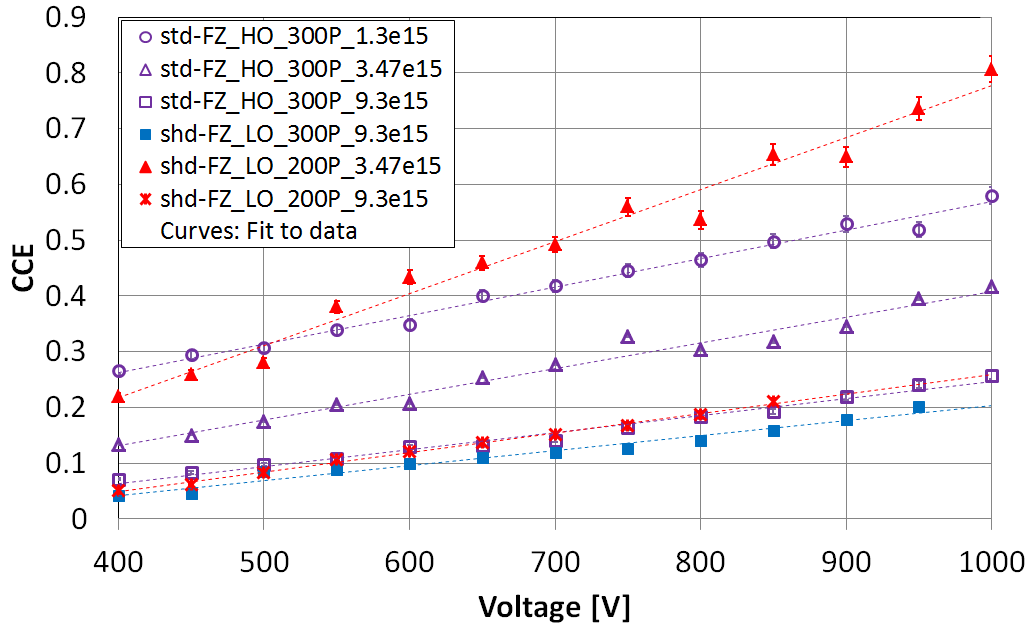}\label{CCE_V23}}\hspace{1mm}%
     \subfloat[ 
     ]{\includegraphics[width=.49\textwidth]{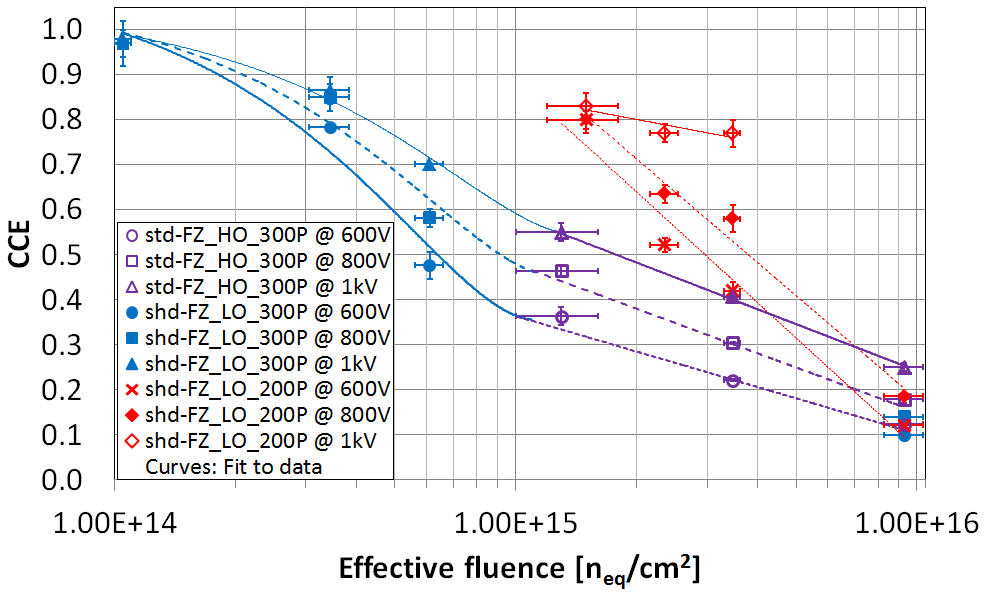}\label{CCE_F23}}\\
     \subfloat[ 
     ]{\includegraphics[width=.50\textwidth]{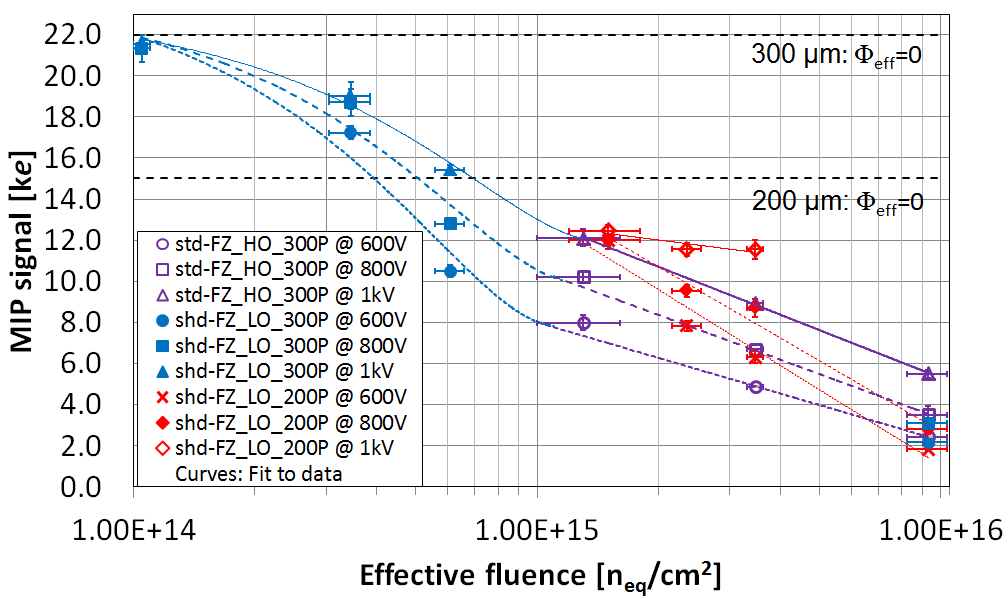}\label{ke_F23}}
    \caption{\small CCE 
as a function of voltage and fluence, as well as $Q_\textrm{coll}$ evolution with fluence for 300 and 200-$\upmu$m thick $n$-on-$p$ sensors at -30 $^{\circ}$C in 
    higher-fluence region than anticipated for the two thicknesses at HGCAL. 
Sensor identification is indicated in the legends, as well as effective fluences (in units of n$_\textrm{eq}$cm$^\textrm{-2}$) in panel (a). 
(a) Measured CCE($V$) with linear fits to data 
included. 
(b) Measured 
CCE($\Phi_\textrm{eff}$) including all measured data points for both sensor thicknesses. The fits to 300P data are polynomial and logarithmic below and above the fluence of $1.3\times10^{15}~\textrm{n}_\textrm{eq}\textrm{cm}^{-2}$, respectively. The fits to 200P data are logarithmic. 
(c) Measured $Q_\textrm{coll}$($\Phi_\textrm{eff}$) from MIP charge deposition with $Q_\textrm{coll}$ for both active thicknesses at CCE = 1 indicated \cite{Bichsel1988,Patri2016}. 
} 
\label{fig_CCE23}
\end{figure*}

The CCE results of the six $n$-on-$p$ samples 
at neutron fluences 
up to 
$10^{16}~\textrm{n}_\textrm{eq}\textrm{cm}^{-2}$ are presented in Figure~\ref{fig_CCE23}. 
Both the linear voltage dependence and the absolute values of CCE of the two highest fluence 300P sensors in Figure~\ref{CCE_V23} are closely in line with 
the results of an earlier charge collection study of 300-$\upmu\textrm{m}$ sensors at extreme neutron fluences \cite{Kramberger2013}.
%
Displayed in Figures~\ref{CCE_F23} and~\ref{ke_F23}, the fluence evolution of the charge collection in 300P sensors 
can be estimated by second-order polynomial and logarithmic fits below and above the fluence of $1.3\times10^{15}~\textrm{n}_\textrm{eq}\textrm{cm}^{-2}$, respectively. After presenting full depletion voltage evolution at fluences above $1\times10^{15}~\textrm{n}_\textrm{eq}\textrm{cm}^{-2}$ in Section~\ref{secVfd}, the observed fluence dependence of charge collection at extreme fluences is further discussed in Section~\ref{Discussion}. 

\begin{figure*}
     \centering
     \includegraphics[width=.70\textwidth]{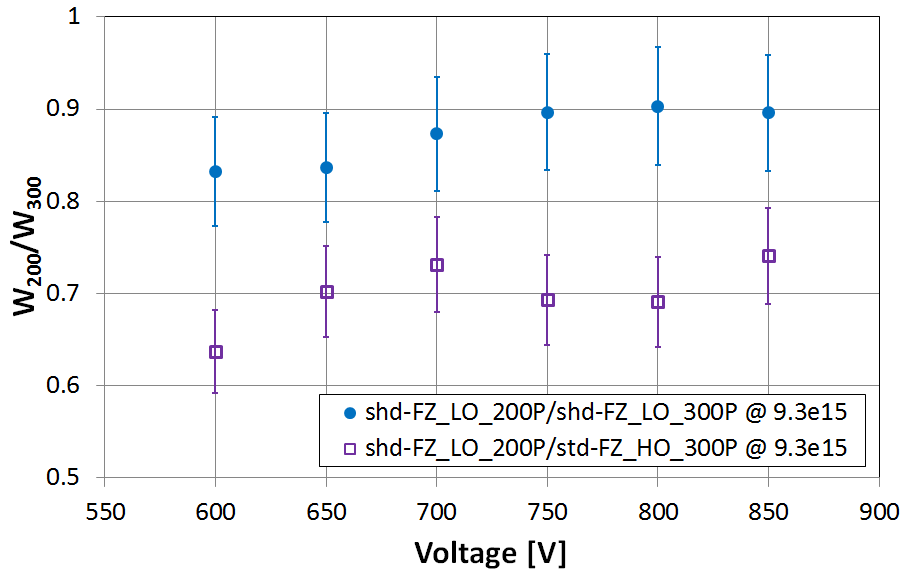}
    \caption{\small 
Depletion-depth ratios as a function of voltage for 
    300-$\upmu$m (300P) and 200-$\upmu$m (200P) thick $n$-on-$p$ diodes 
    at the highest fluence in Figures~\ref{CCE_F23} and~\ref{ke_F23}, and calculated by Eq.~\ref{eq4}. 
Sensor identification and effective fluences (in units of n$_\textrm{eq}$cm$^\textrm{-2}$) are indicated in the legend. 
}
\label{W23_1e16}
\end{figure*}

\begin{linenomath}
Additionally, the advantageous CCE performance of the 200P sensor in the range $(1.5-3.5)\times10^{15}~\textrm{n}_\textrm{eq}\textrm{cm}^{-2}$ ($\sim$20\%, 30\%, and 36\% higher at $3.5\times10^{15}~\textrm{n}_\textrm{eq}\textrm{cm}^{-2}$ compared to 300P for 600 V, 800 V, 
and 1 kV, 
respectively) diminishes almost completely at the highest fluence in Figure~\ref{CCE_F23}. This is also observed 
throughout the voltage range of Figure~\ref{CCE_V23}. Similar convergence between sensor thicknesses around $1\times10^{16}~\textrm{n}_\textrm{eq}\textrm{cm}^{-2}$ was reported in an earlier study \cite{Affolder2011b}. 
The highest fluence CCE and $Q_\textrm{coll}$ values in Figures~\ref{CCE_F23} and~\ref{ke_F23} are presented in Table~\ref{table_cc23}. 
The results display some differences in 
$Q_\textrm{coll}$ between sensors with high (HO) and low (LO) bulk oxygen content. 
%
To investigate further the 
similar CCE performance of the 200P and 300P sensors at highest neutron fluence, 
the following relations were considered.
The 
CCE of an irradiated sensor 
is formulated as a product of two terms, geometrical factor ($\textrm{CCE}_\textrm{GF}$) and trapping factor ($\textrm{CCE}_\textrm{t}$) \cite{RD39_2006,Kraner1993} 
%
\begin{equation}\label{eq2}
\textrm{CCE}=\textrm{CCE}_\textrm{GF}\times\textrm{CCE}_\textrm{t}=\frac{W}{d}\frac{\tau_\textrm{t}}{t_\textrm{dr}}(1-e^{-\frac{t_\textrm{dr}}{\tau_\textrm{t}}}),
\end{equation}
where $W$ is the depletion depth, $d$ the active thickness of the sensor, $\tau_\textrm{t}$ the trapping time constant for electrons or holes, and $t_\textrm{dr}$ the carrier drift time in the depletion region. To solve for $W$ from the CCE data would first require to determine $\tau_\textrm{t}$. This would demand a 
short-range charge injection (e.g. red laser), where only one type of charge carriers generate the transient signal. 
%
%
%
However, when it is considered that the investigated samples were exposed to equal fluence (leading to identical $\tau_\textrm{t}$), the CCE results were acquired at equal 
bias voltage $V$, and that the observed carrier drift times were close to equal (mean $t_\textrm{dr}$ of the 200P and the two 300P diodes at the highest fluence in Figure~\ref{CCE_F23} was $3.7\pm0.2$ ns), 
leads to the relationship  
%
\begin{equation}\label{eq4}
\frac{W_\textrm{200P}}{W_\textrm{300P}}=\frac{\textrm{CCE}(V)_\textrm{200P}}{\textrm{CCE}(V)_\textrm{300P}}\frac{d_\textrm{200P}}{d_\textrm{300P}}. 
\end{equation}
Figure~\ref{W23_1e16} shows the depletion-depth ratios when the CCE($V$) results from Figure~\ref{CCE_V23} are inserted into Eq.~\ref{eq4}. The ratios remain close to constant in the investigated voltage range with the average 
values $W_{\textrm{200P}_{-}\textrm{LO}}/W_{\textrm{300P}_{-}\textrm{LO}}=0.87\pm0.15$ and $W_{\textrm{200P}_{-}\textrm{LO}}/W_{\textrm{300P}_{-}\textrm{HO}}=0.70\pm0.12$. Thus, for low oxygen concentration 
200P and 300P sensors the result indicates that the charge is collected from equal depth in both sensors within uncertainty. However, the ratio is significantly smaller between low and high 
oxygen concentration 200P and 300P sensors, respectively. This could suggest a limited beneficial influence of the higher oxygen concentration on the build-up of negative space charge 
at extreme neutron fluence (as opposed to no effect at about 30-fold lower neutron fluences in \cite{Lindstrom2003,Kramberger2002p}), leading to larger depletion depth at given voltage. Trapping times have not been observed to be significantly influenced by the oxygen content after hadron irradiation \cite{Kramberger2002p}. Thus, Eq.~\ref{eq4} is expected to be also applicable for comparison between LO and HO sensors.
\end{linenomath}
%
\begin{table*}[!t]
\centering
\caption{CCE and $Q_\textrm{coll}$ of 200P and 300P sensors in Figures~\ref{CCE_F23} and~\ref{ke_F23}, respectively, at the fluence of $(9.3\pm1.1)\times10^{15}~\textrm{n}_\textrm{eq}\textrm{cm}^{-2}$ and for 600 and 800 V. 
} 
\label{table_cc23}
\begin{tabular}{|c|c|c|c|c|}
\hline
\multirow{2}{*}[0.5em]{{\bf Sensor thickness \&}} & \multirow{2}{*}[0.5em]{{\bf CCE(600 V)}} & \multirow{2}{*}[0.5em]{{\bf $Q_\textrm{coll}$(600 V)}} 
& \multirow{2}{*}[0.5em]{{\bf CCE(800 V)}} & \multirow{2}{*}[0.5em]{{\bf $Q_\textrm{coll}$(800 V)}}\\
{\bf type} & & [k$e$] & & [k$e$]\\
\hline
shd-FZ$_{-}$LO$_{-}$200P & $0.121\pm0.003$ & $1.82\pm0.05$ & $0.186\pm0.005$ & $2.79\pm0.08$\\
\hline
shd-FZ$_{-}$LO$_{-}$300P & $0.099\pm0.007$ & $2.18\pm0.15$ & $0.140\pm0.005$ & $3.08\pm0.11$\\
\hline
std-FZ$_{-}$HO$_{-}$300P & $0.111\pm0.006$ & $2.44\pm0.12$ & $0.16\pm0.02$  & $3.5\pm0.4$\\
\hline
\end{tabular}
\end{table*}
%
\newpage
\subsection{Full depletion voltages}
\label{secVfd}
\begin{figure*}
     \centering
     \subfloat[ 
     ]{\includegraphics[width=.5\textwidth]{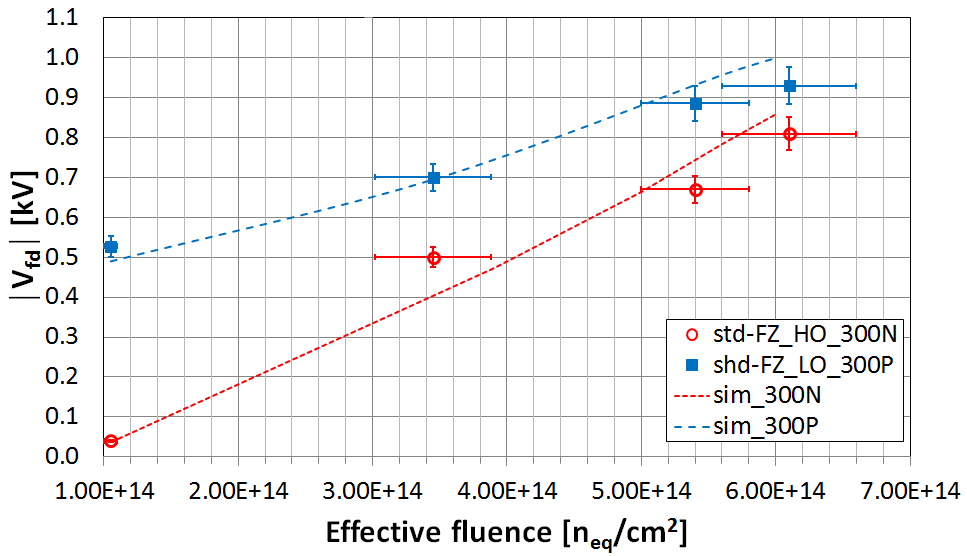}\label{Vfd300}}\hspace{1mm}%
     \subfloat[ 
     ]{\includegraphics[width=.49\textwidth]{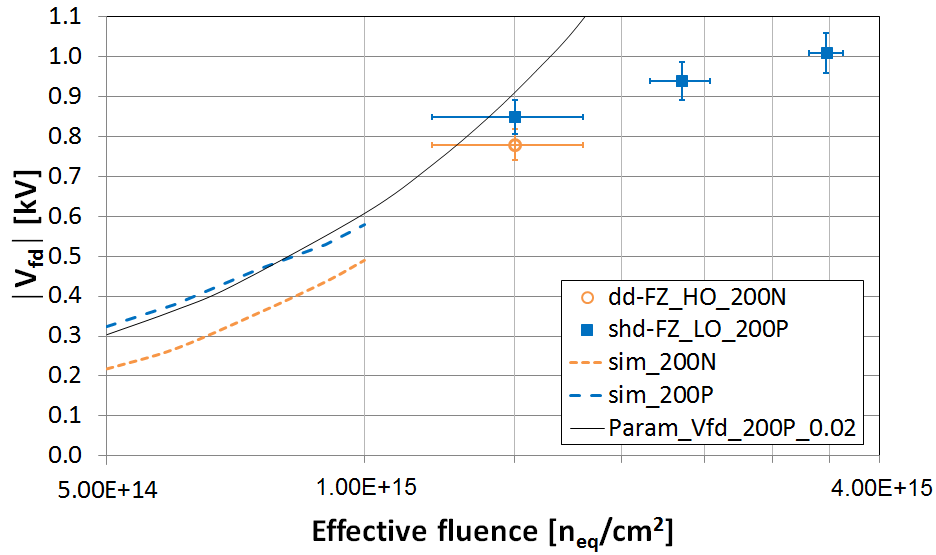}\label{Vfd200}}\\
     \subfloat[ 
     ]{\includegraphics[width=.50\textwidth]{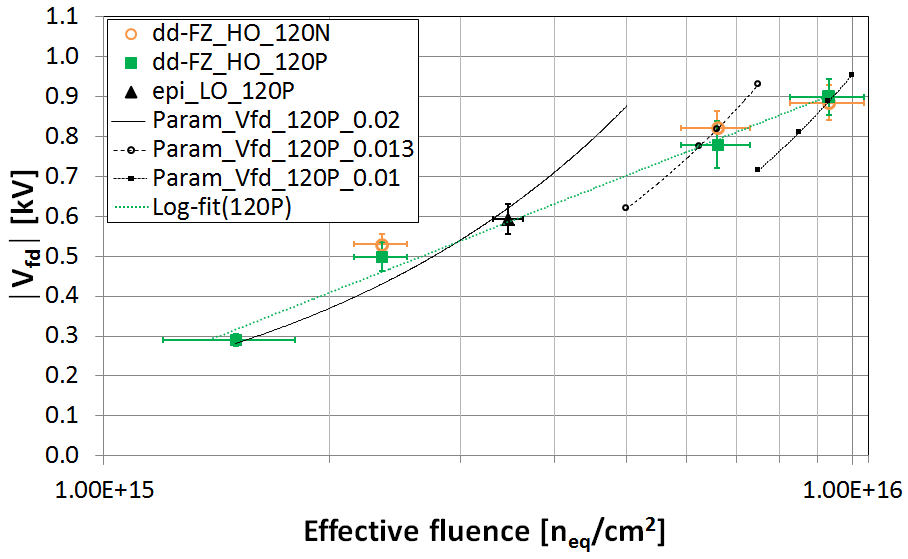}\label{Vfd120}}
    \caption{\small 
    Evolution of full depletion voltage as a function of $\Phi_\textrm{eff}$ for the fluence ranges anticipated for the three sensor thicknesses at HGCAL. Simulations are applied up to the neutron defect model's validated fluence limit \cite{Eber2013}. 
The corresponding $CV$-curves are presented in 
Figures~\ref{fig_CV300} and~\ref{fig_CV200_120}. (a) Measured (markers) and simulated (dashed curves) $V_{\textrm{fd}}(\Phi_\textrm{eff})$ for 
300-$\upmu$m thick sensors. Initial $V_{\textrm{fd}}$ before irradiation for 300N and 300P were 200 V and 300 V, respectively.  
(b) Measured and simulated $V_{\textrm{fd}}(\Phi_\textrm{eff})$ for 
200-$\upmu$m thick sensors. $V_{\textrm{fd}}$ before irradiation for 200N and 200P were 95 V and 130 V, respectively. Parametrized $V_{\textrm{fd}}(\Phi_\textrm{eff})$ was calculated using Eq.~\ref{eq2b} with standard value of $g_\textrm{eff}=0.02~\textrm{cm}^{-1}$ \cite{Cartiglia2018}.
(c) Measured $V_{\textrm{fd}}(\Phi_\textrm{eff})$ for 
120-$\upmu$m thick sensors. $V_{\textrm{fd}}$ before irradiation for both 120N and 120P was 50 V, while for epitaxial 120P it was 20 V. Parametrized $V_{\textrm{fd}}(\Phi_\textrm{eff})$ was calculated using Eq.~\ref{eq2b} with standard ($0.02~\textrm{cm}^{-1}$), intermediate ($0.013~\textrm{cm}^{-1}$) and saturated ($0.01~\textrm{cm}^{-1}$ \cite{Cartiglia2018}) values of $g_\textrm{eff}$. The logarithmic fit value for 120P sensors is $(0.33\pm0.11)\ln(x)-(11\pm4)$. 
}
\label{fig_Vfd}
\end{figure*}

The fluence evolution of full depletion voltage ($V_\textrm{fd}$) is presented in Figure~\ref{fig_Vfd} with measured and simulated results of 300- and 200-$\upmu$m-thick diodes in Figures~\ref{Vfd300} and~\ref{Vfd200}, respectively, and measured results of 
120-$\upmu$m-thick diodes in Figure~\ref{Vfd120}. 
The simulated results in the 
lower-fluence region 
were produced by applying 
the neutron defect model 
using as an input the leakage current extracted $\Phi_\textrm{eff}$ from Table~\ref{table_fluence}, and frequencies in the range of the measured $CV$ ($\mathcal{O}(10^{4}~\textrm{Hz})$). 
The simulated $V_\textrm{fd}$ was extracted both from $CV$-curve and from the voltage where electric field had extended throughout the active thickness of the sensor. The frequency was tuned until the two results were in agreement. 

\begin{linenomath}
$V_\textrm{fd}$ was defined as the crossing point of the two linear fits made on the dynamic and static regions of the reciprocal $C^2$ curve, as demonstrated 
in Figures~\ref{fig_CV300} and~\ref{fig_CV200_120}. 
Measurements were carried out at frequencies from 20 kHz to 50 kHz, 
and to take into account the increasing frequency dependence \cite{Li1991,Riedler1998} of $V_\textrm{fd}$ with fluence, the two verifying methods utilized were the aforementioned simulations up to $1\times10^{15}~\textrm{n}_\textrm{eq}\textrm{cm}^{-2}$ for both sensor polarities and analytical method using parametrized effective bulk doping for $n$-on-$p$ sensors \cite{Cartiglia2018,Terada1996,Sadro2017}. To reach analytical $V_{\textrm{fd}}(\Phi_\textrm{eff})$, the fluence evolution of effective bulk doping was considered as 
%
\begin{equation}\label{eq2a}
N_\textrm{A}(\Phi_\textrm{eff})=g_\textrm{eff}\Phi_\textrm{eff}+N_\textrm{A}(0)e^{-c\Phi_\textrm{eff}},
\end{equation}
where $g_\textrm{eff}$ is the acceptor creation coefficient with a standard value of $0.02~\textrm{cm}^{-1}$ and saturated value of $0.01~\textrm{cm}^{-1}$ at high fluences \cite{Cartiglia2018,Balbuena2009}. $N_\textrm{A}(0)$ and $N_\textrm{A}(\Phi_\textrm{eff})$ are the acceptor concentrations before and after irradiation, respectively, and $c=3\times10^{-14}~\textrm{cm}^2$ \cite{Sadro2017} is a constant that depends on the initial acceptor concentration and on the type of irradiation. $V_{\textrm{fd}}(\Phi_\textrm{eff})$ is then given by the relation \cite{Sze1981}
\begin{equation}\label{eq2b}
V_\textrm{fd}(\Phi_\textrm{eff})=\frac{ed^2}{2\epsilon_\textrm{s}}N_\textrm{A}(\Phi_\textrm{eff}),
\end{equation}
where $e$ is the elemental charge, $d$ the active thickness of the sensor and $\epsilon_\textrm{s}$ the permittivity of silicon. Values for $N_\textrm{A}(0)$ in Figures~\ref{Vfd200} and~\ref{Vfd120} were $4.03\times10^{12}~\textrm{cm}^{-3}$ for shd-FZ 200P-sensors, and $4.19\times10^{12}$ and $2.1\times10^{12}~\textrm{cm}^{-3}$ for dd-FZ 120P- and epitaxial 120P-sensors, respectively. 
\end{linenomath}

Since full depletion can be monitored in the simulation by the extension of the electric field throughout the active thickness of the sensor, it can be used to verify measured $V_\textrm{fd}$-values. 
Displayed in Figure~\ref{Vfd300}, the simulation is within 
uncertainties of all data points and exhibits linear increase with fluence. 

When analytical $V_{\textrm{fd}}(\Phi_\textrm{eff})$ with standard $g_\textrm{eff}$ is compared to the simulated and measured values of 200P-sensors in Figure~\ref{Vfd200}, the results are in close agreement up to the fluence of $1.5\times10^{15}~\textrm{n}_\textrm{eq}\textrm{cm}^{-2}$. The corresponding comparison for 120P-sensors in Figure~\ref{Vfd120} shows agreement up to the fluence of $(3.47\pm0.16)\times10^{15}~\textrm{n}_\textrm{eq}\textrm{cm}^{-2}$. In addition, when the measured $V_{\textrm{fd}}$ is estimated to be about 1.5 kV for the 300P sensor 
at the lowest fluence of $(1.3\pm0.3)\times10^{15}~\textrm{n}_\textrm{eq}\textrm{cm}^{-2}$ in Figure~\ref{Vfd23_1e16}, this is reproduced by the analytical $V_{\textrm{fd}}(\Phi_\textrm{eff})$ at the fluence of $1.1\times10^{15}~\textrm{n}_\textrm{eq}\textrm{cm}^{-2}$. 
Beyond these fluences the measured $V_{\textrm{fd}}$ results display less steep increase with fluence than the analytical model with standard $g_\textrm{eff}$. However, analytical $V_{\textrm{fd}}(\Phi_\textrm{eff})$ with intermediate ($0.013~\textrm{cm}^{-1}$) and saturated $g_\textrm{eff}$-values reproduces closely the data points at second highest and highest fluences in Figure~\ref{Vfd120}, respectively. 
Additionally, the fluence evolution of $V_\textrm{fd}$ above $1\times10^{15}~\textrm{n}_\textrm{eq}\textrm{cm}^{-2}$ in Figure~\ref{Vfd120} is logarithmic, with a fit value of $(0.33\pm0.11)\ln(x)-(11\pm4)$, as opposed to the linear behavior below $1\times10^{15}~\textrm{n}_\textrm{eq}\textrm{cm}^{-2}$ seen in Figure~\ref{Vfd300}.  
These observations 
support the saturating behavior of the acceptor creation mechanism at high fluences, 
as described in ref. \cite{Cartiglia2018}.
%
%

Also visible in Figure~\ref{Vfd300} is the effect of 
SCSI, discussed in Section~\ref{CCEresults}, in the $n$-type doped silicon substrate, which results in the decrease of $V_\textrm{fd}$(300N) from $200~\textrm{V}$ before irradiation to about 40 V at $(1.05\pm0.05)\times10^{14}~\textrm{n}_\textrm{eq}\textrm{cm}^{-2}$. Thus, 
the combined information from measured and simulated data suggests that 300N sensors operated at 600 V 
are fully depleted below the fluence of $5\times10^{14}~\textrm{n}_\textrm{eq}\textrm{cm}^{-2}$, 
while the corresponding fluence limit for 300P sensors is below $3\times10^{14}~\textrm{n}_\textrm{eq}\textrm{cm}^{-2}$. Furthermore, the results 
indicate that operating at 800 V 
extends the fluence region where the 300P sensor is still fully depleted 
to about $4\times10^{14}~\textrm{n}_\textrm{eq}\textrm{cm}^{-2}$.

Figure~\ref{Vfd200} 
shows that both polarities of 200-$\upmu\textrm{m}$-thick sensors 
are 
fully depleted at 800 V 
below a fluence of $(1.5\pm0.3)\times10^{15}~\textrm{n}_\textrm{eq}\textrm{cm}^{-2}$ and below about $1\times10^{15}~\textrm{n}_\textrm{eq}\textrm{cm}^{-2}$ 
at 600 V. 
At $(3.47\pm0.16)\times10^{15}~\textrm{n}_\textrm{eq}\textrm{cm}^{-2}$, the 120P epitaxial diode is still reaching full depletion at 600 V in Figure~\ref{Vfd120}, explaining the high CCE performance in e.g. Figure~\ref{CCE_F120}. 
%

\section{Discussion}
\label{Discussion}
%
\begin{enumerate}
\item The 
300N std-FZ(HO) sensor  
%
displays better CCE performance to 300P shd-FZ(LO) for a given operating voltage above about $4\times10^{14}~\textrm{n}_\textrm{eq}\textrm{cm}^{-2}$ due to 
SCSI, which enables it to be fully depleted at significantly lower voltages than the 300P sensor (the 300N sensor is fully depleted at 600 V 
up to the fluence of about $5\times10^{14}~\textrm{n}_\textrm{eq}\textrm{cm}^{-2}$, while the 300P sensor is not fully depleted at 800 V 
above the fluence of about $4\times10^{14}~\textrm{n}_\textrm{eq}\textrm{cm}^{-2}$). At the fluence of $(6.1\pm0.5)\times10^{14}~\textrm{n}_\textrm{eq}\textrm{cm}^{-2}$ $\Delta$CCE of $(17\pm2)\%$ in favor of the 300N sensor is observed at 
600 and 800 V, and 12\% at 1 kV. Similar CCE behavior was observed 
in the earlier 
studies \cite{Phase2,Curras2017}.
\item No clear difference was observed in the CCE performance of 120-$\upmu\textrm{m}$-thick dd-FZ(HO) diodes between the sensor 
polarities at the highest 
fluence, which is 
also 
in 
agreement with 
the earlier observations \cite{Phase2,Curras2017}. 
One of the main advantages of the $n$-on-$p$ sensor 
at fluences above $1\times10^{15}~\textrm{n}_\textrm{eq}\textrm{cm}^{-2}$ is the superposition of the weighting field and the electric field maxima on the segmented side of the planar sensor, while in the $p$-on-$n$ sensor after SCSI, these are located on the opposite sides of a planar sensor. In strip and pixel sensors used in tracking and vertexing detectors, the effect of the weighting field is optimized by sensor geometry, which favors electron contribution to the signal collected at the n$^+$ electrode \cite{Kramberger2013,Kramb02}, while in the large area pad detectors, 
the benefit from the weighting field 
(simply the inverse of the sensor thickness 1/$d$ \cite{Ramo1939}) is minimal, 
leading to negligible differences in CCE performances between sensor 
types at high fluences. 
\item The slowing down of the degradation of charge collection of 300P-sensors at fluences above $1\times10^{15}~\textrm{n}_\textrm{eq}\textrm{cm}^{-2}$ in Figures~\ref{CCE_F23} and~\ref{ke_F23} is linearly correlated to the slowing of $V_\textrm{fd}$ increase in Figure~\ref{Vfd120}, with slopes of $-(0.42\pm0.06)~\textrm{kV}^{-1}$ for the three voltages in Figures~\ref{CCE_F23} and~\ref{ke_F23}. This suggests beneficial influence to the charge collection performance from the saturation of acceptor creation at extreme fluences. 
%
\item TCAD simulation tuning with 
input from $CV/IV$-measurements before irradiation and IR-TCT parameters resulted in closely reproduced transient signal shapes that minimize error sources for 
modeling of radiation damage in silicon. 
Further study will need to be conducted to develop a neutron defect model that is able to reproduce all properties of irradiated silicon sensors observed in this 
and previous studies in the fluence range of $(0.1-1)\times10^{16}~\textrm{n}_\textrm{eq}\textrm{cm}^{-2}$.
\end{enumerate}

\section{Summary}
\label{Conclusions}
After neutron irradiation campaigns at RINSC and MNRC, 
the electrical characterization 
of 
12 and 18 irradiated samples diced from 8-inch and 6-inch wafers, respectively, was completed and the most probable effective fluences were extracted from the leakage currents above $V_\textrm{fd}$. 
The results of the CCE study 
of test diodes at -30 $^{\circ}$C and for operating voltages of 600, 800, and 1000 V, 
along with the results of the electrical properties, indicate the following: 
%
%
\begin{itemize}
\item For 300-$\upmu\textrm{m}$-
thick sensors: 
\begin{itemize}
\item TCAD simulations 
are in close agreement and 
complement the CCE results 
from the IR-TCT measurements.
\item CCEs of both sensor polarities gain about 10\% 
by increasing the operating voltage from 600 V to 800 V 
at highest expected lifetime fluences ($\sim6\times10^{14}~\textrm{n}_\textrm{eq}\textrm{cm}^{-2}$), resulting in CCEs of $\sim$76\% and 60\% for the 300N 
and 300P 
sensors, respectively. Corresponding gains at 1 kV are 16\% and 22\%, respectively. At the lowest lifetime fluence ($1\times10^{14}~\textrm{n}_\textrm{eq}\textrm{cm}^{-2}$), 
a CCE of $\sim$96\% was observed for both sensor types. 
The 300N sensor 
displays better CCE performance compared to the 300P 
above the fluence of about $4\times10^{14}~\textrm{n}_\textrm{eq}\textrm{cm}^{-2}$ due to 
SCSI. 
%
\item In the fluence range $1\times10^{15}-1\times10^{16}~\textrm{n}_\textrm{eq}\textrm{cm}^{-2}$ 
the observed slowing down of the degradation of charge collection performance is linearly correlated to the slowing of $V_\textrm{fd}$ increase, suggesting beneficial influence from the saturation of acceptor creation at extreme fluences. 
\end{itemize}
\item For 200-$\upmu\textrm{m}$-
thick sensors: 
\begin{itemize}
\item The CCE of the 200P 
increases by $\sim$11\% and 25\% 
from operating at 800 V and 1 kV, respectively, instead of 600 V, at 
the fluence of $(2.35\pm0.19)\times10^{15}~\textrm{n}_\textrm{eq}\textrm{cm}^{-2}$ ($\approx\Phi_\textrm{max}$), resulting in CCEs of $\sim$63\% and 77\%, respectively. At the lowest lifetime 
fluence ($5\times10^{14}~\textrm{n}_\textrm{eq}\textrm{cm}^{-2}$), 
a CCE of $\sim$91\% 
is expected from simulation for both sensor types. 
Furthermore, the CCE performance of 200P sensor displays 16\% and 35\% improvement 
from increasing the operating voltage from 600 V to 
800 V and 1 kV, respectively, at the fluence of $(3.47\pm0.16)\times10^{15}~\textrm{n}_\textrm{eq}\textrm{cm}^{-2}$, resulting in CCEs of $\sim$60\% and 77\%, respectively.
\item Combined measured, simulated and analytically calculated $V_\textrm{fd}$ results suggest that the 200P-sensor is not fully depleted at 800 V 
beyond the fluence of $(1.3\pm0.3)\times10^{15}~\textrm{n}_\textrm{eq}\textrm{cm}^{-2}$, while 
it can be operated at 600 V fully depleted below the 
fluence of $1\times10^{15}~\textrm{n}_\textrm{eq}\textrm{cm}^{-2}$.
\end{itemize}
\item For 120-$\upmu\textrm{m}$-
thick sensors:
\begin{itemize}
\item The collected charges from a MIP-induced charge deposition 
close to the fluence of $10^{16}~\textrm{n}_\textrm{eq}\textrm{cm}^{-2}$ show values of 0.58 at 600 V, 0.87 at 800 V, and 1.17 fC at 1 kV operating voltage. Thus, the sensors require more than 800 V biasing 
to collect at least 1 fC from a single-MIP-charge-injection at highest lifetime fluences ($\sim1\times10^{16}~\textrm{n}_\textrm{eq}\textrm{cm}^{-2}$). 
\item The CCE performance of both polarity sensors improves by 20\% and 40\% 
by increasing the operating from 600 V to 
800 V and 1 kV, respectively, at the fluence of $(9.3\pm1.1)\times10^{15}~\textrm{n}_\textrm{eq}\textrm{cm}^{-2}$. 
At lowest 
investigated fluences ($(1.5\pm0.3)\times10^{15}~\textrm{n}_\textrm{eq}\textrm{cm}^{-2}$), 
both sensor types display essentially non-degraded CCE performance.
\item 
Similar CCE performance 
was observed 
between sensor 
types at the highest 
fluence due to minimal benefit from the weighting field to electron collection in $n$-on-$p$ pad sensors. 
\item Combined measured and analytically calculated $V_\textrm{fd}$ results suggest that 
120P-sensors are not fully depleted at 800 V 
above the fluence of about $6\times10^{15}~\textrm{n}_\textrm{eq}\textrm{cm}^{-2}$, while 
sensors of both polarities are fully depleted at 
600 V below the fluence of $(3.47\pm0.16)\times10^{15}~\textrm{n}_\textrm{eq}\textrm{cm}^{-2}$. Fluence evolution of $V_\textrm{fd}$ above $1\times10^{15}~\textrm{n}_\textrm{eq}\textrm{cm}^{-2}$ appears logarithmic and shows evidence of the saturation of acceptor creation mechanism above $5\times10^{15}~\textrm{n}_\textrm{eq}\textrm{cm}^{-2}$.
%
\end{itemize}
\item 
CCE analysis suggests that the similar charge collection performance between 200P(LO) and 300P(LO) sensors at the fluence of $(9.3\pm1.1)\times10^{15}~\textrm{n}_\textrm{eq}\textrm{cm}^{-2}$ 
results from close-to-equal depletion regions. Higher $Q_\textrm{coll}$ in 300P(HO) relative to 200P/300P(LO) at equal fluence 
could be evidence of small benefit to radiation hardness from higher oxygen content in the sensor bulk at extreme neutron fluences. Further studies will be needed to verify this observation. 
\end{itemize}

\section*{Acknowledgements}
We thank K. Zinsmeyer, P. Cruzan and C. Perez of TTU for their expert technical support, as well as J. Christian and P. Wibert for their contributions to the sensor characterization measurements. 
%

\bibliography{mybibfile}

\newpage
\begin{appendices}
\label{Appendix}
\section{Appendix: $CV$-results}
%
\begin{figure*}[htb]
     \centering
     \subfloat[ 
     ]{\includegraphics[width=.45\textwidth]{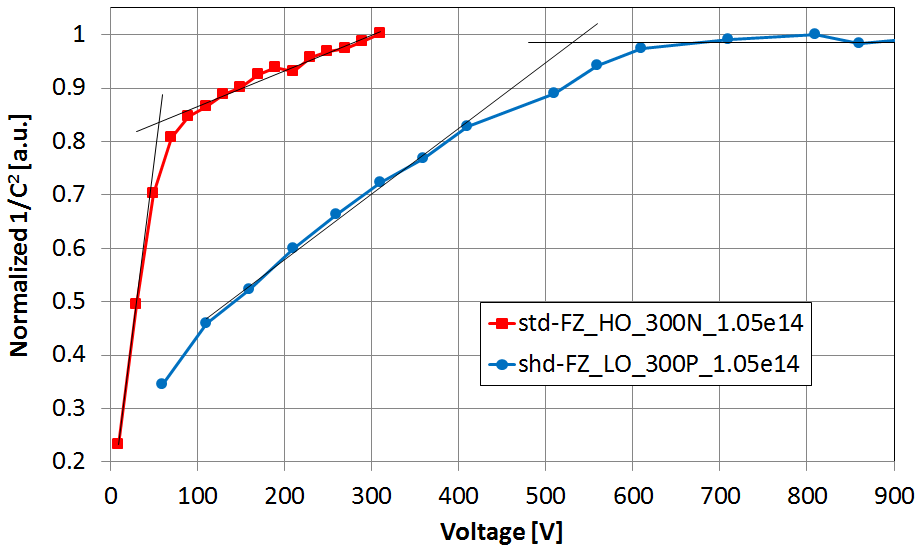}\label{CV300NP}}\hspace{1mm}%
     \subfloat[ 
     ]{\includegraphics[width=.49\textwidth]{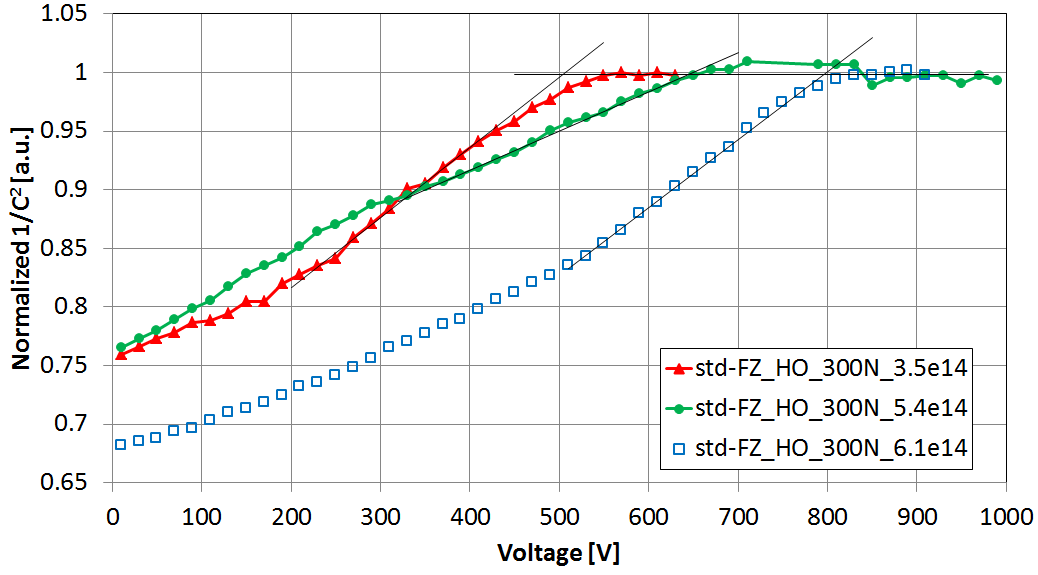}\label{CV300N}}\\
     \subfloat[ 
     ]{\includegraphics[width=.495\textwidth]{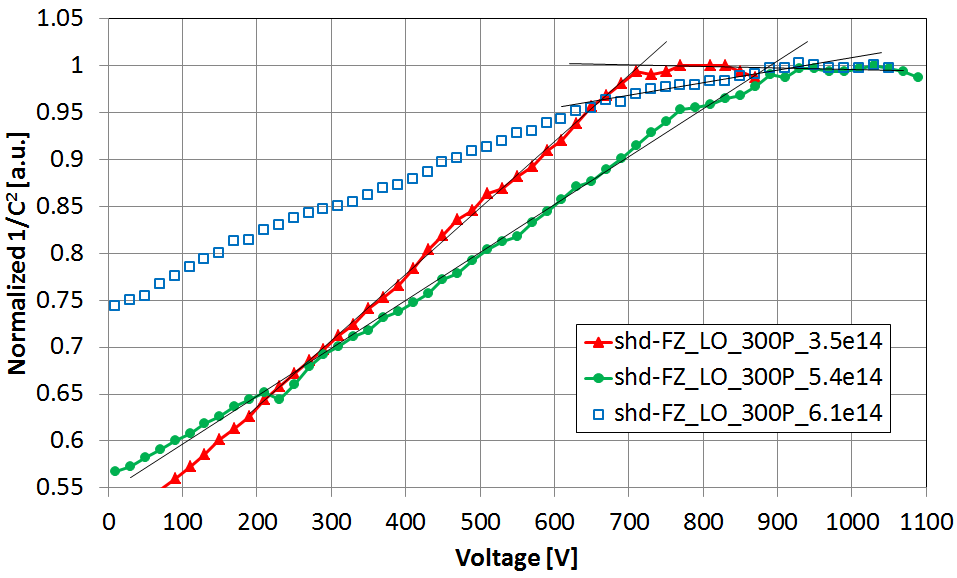}\label{CV300P}}\hspace{1mm}%
      \subfloat[ ]{\includegraphics[width=.498\textwidth]{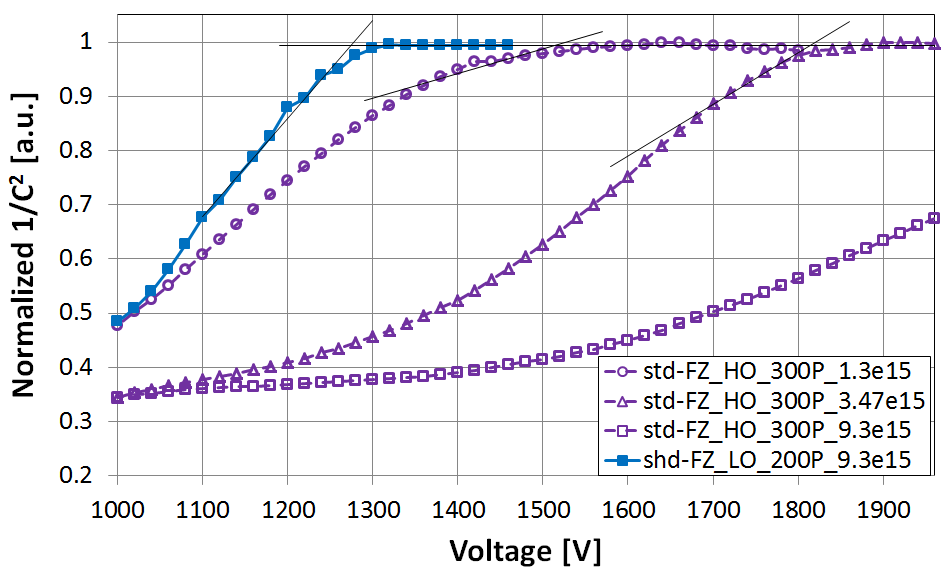}\label{Vfd23_1e16}}
    \caption{\small Measured reciprocal $C^2$ as a function of voltage and extracted $V_{\textrm{fd}}$-values (crossing point of two linear fits) for 
    (a) 300N/P sensors at $(1.05\pm0.05)\times10^{14}~\textrm{n}_\textrm{eq}\textrm{cm}^{-2}$, (b) 300N sensors and (c) 300P sensors corresponding to Figure~\ref{Vfd300}. Also shown in (d) are the results for 300-$\upmu$m (300P) and 200-$\upmu$m (200P) thick $n$-on-$p$ diodes irradiated up to 
    about $10^{16}~\textrm{n}_\textrm{eq}\textrm{cm}^{-2}$.  
    Sensor identification and effective fluences (in units of n$_\textrm{eq}$cm$^\textrm{-2}$) 
are indicated in the legends. 
}
\label{fig_CV300}
\end{figure*}
\begin{figure*}[htb]
     \centering
     \subfloat[ 
     ]{\includegraphics[width=.49\textwidth]{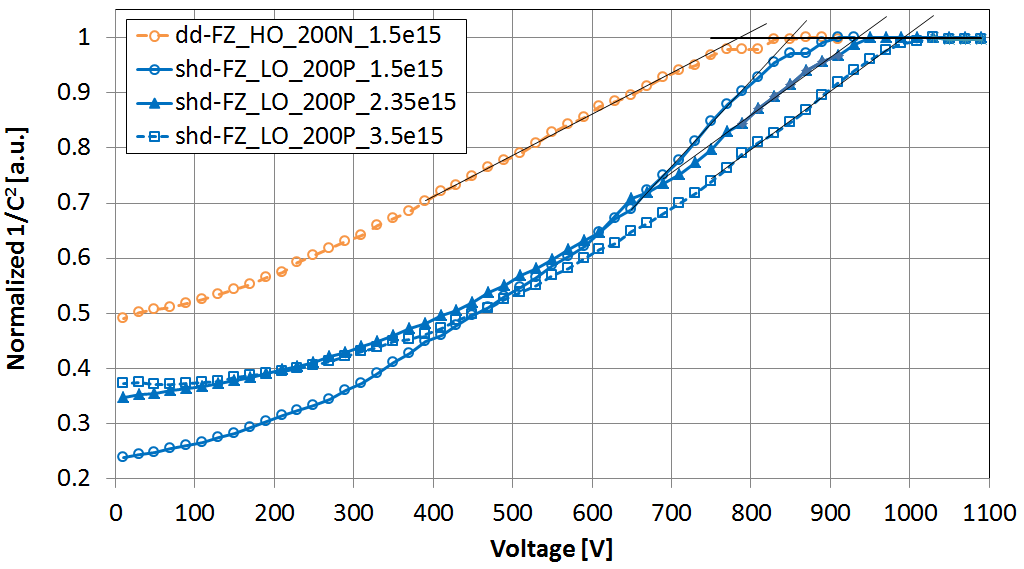}\label{CV200}}\hspace{1mm}%
     \subfloat[ 
     ]{\includegraphics[width=.49\textwidth]{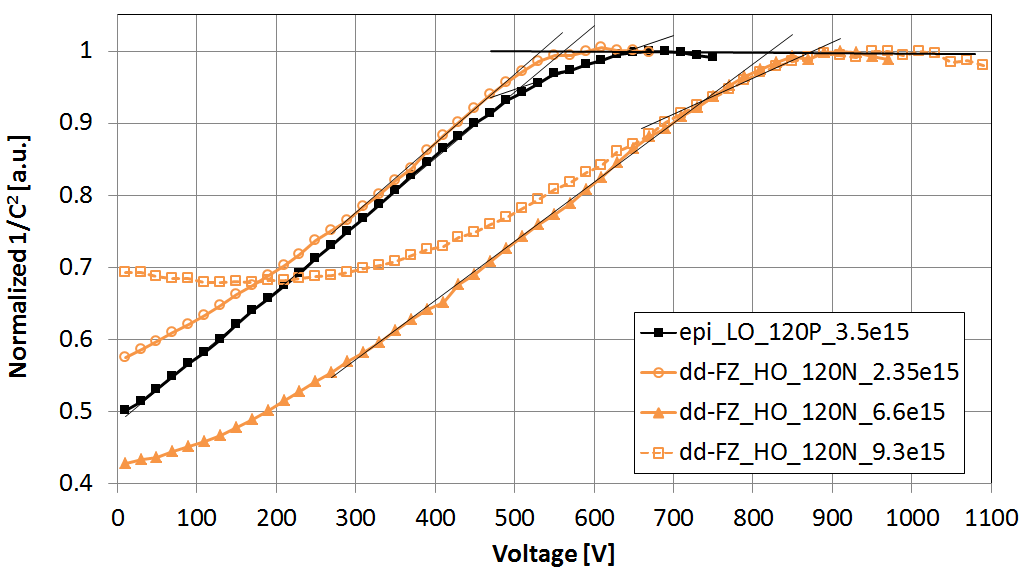}\label{CV120N}}\\
     \subfloat[ 
     ]{\includegraphics[width=.50\textwidth]{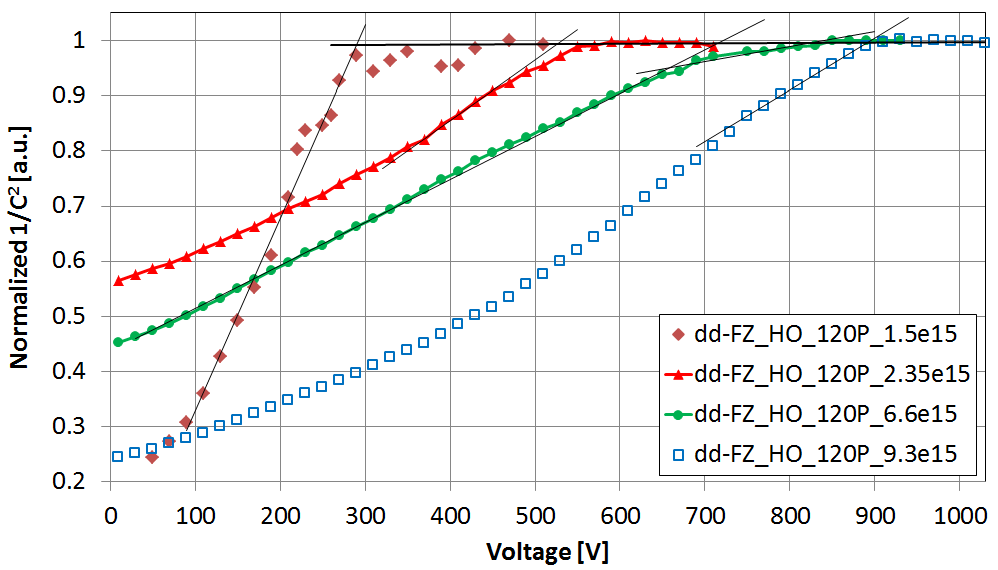}\label{CV120P}}
    \caption{\small Measured reciprocal $C^2$ as a function of voltage and extracted $V_{\textrm{fd}}$-values (crossing point of two linear fits) for 
    (a) 200N/P sensors, (b) epitaxial 120P and 120N sensors, and (c) 120P sensors corresponding to Figures~\ref{Vfd200} and~\ref{Vfd120}, respectively. 
    Sensor identification and effective fluences (in units of n$_\textrm{eq}$cm$^\textrm{-2}$) 
are indicated in the legends. 
}
\label{fig_CV200_120}
\end{figure*}
%

\end{appendices}

\end{document}